\documentclass[a4paper,11pt]{article}

\usepackage{jcappub} 

\usepackage[T1]{fontenc} 

\usepackage{graphicx}

\newcommand{\be}{\begin{equation}}
\newcommand{\ee}{\end{equation}}
\newcommand{\ba}{\begin{eqnarray}}
\newcommand{\ea}{\end{eqnarray}}

\newcommand{\bea}{\begin{eqnarray*}}
\newcommand{\eea}{\end{eqnarray*}}

\newcommand{\cM}{{\cal M}}

\newcommand{\tg}{{\tilde g}}
\newcommand{\tM}{{\tilde M}}
\newcommand{\tR}{{\tilde R}}
\newcommand{\ts}{{\tilde s}}
\newcommand{\pl}{{\partial}}

\title{The effective field theory of K-mouflage}

\author[a]{Philippe Brax,}
\author[a]{Patrick Valageas}

\affiliation[a]{Institut de Physique Th\'eorique, Universit\'e Paris-Saclay, CEA,CNRS,\\
F-91191Gif sur Yvette, France}

\emailAdd{philippe.brax@cea.fr}
\emailAdd{patrick.valageas@cea.fr}

\abstract{We describe K-mouflage models of modified gravity using the effective field
theory of dark energy. We show how the Lagrangian density $K$ defining the K-mouflage
models appears in the effective field theory framework, at both the exact fully nonlinear
level and at the quadratic order of the effective action.
We find that K-mouflage scenarios only generate the operator $(\delta g^{00}_{(u)})^n$
at each order $n$.
We also reverse engineer K-mouflage models by reconstructing the whole effective field
theory, and the full cosmological behaviour, from two functions of the Jordan-frame scale
factor in a tomographic manner.  This parameterisation is directly related to the
implementation of the K-mouflage screening mechanism: screening occurs when $ K'$ is
large in a dense environment such as the deep matter and radiation eras. In this way,
K-mouflage  can be easily implemented as a calculable subclass of models described
by the effective field theory of dark energy  which could be probed by future surveys.}

\begin{document}
\maketitle
\flushbottom

\section{Introduction}

Future surveys such as Euclid \cite{Amendola:2012ys} will actively look for deviations from the $\Lambda$-CDM paradigm. They could be resulting from dynamical properties of dark energy or induced effects of modified gravity on the cosmological background evolution and/or the growth of structure at the linear level\cite{Joyce:2014kja,Koyama:2015vza}. Recently a new approach has been put forward which captures the main features of both approaches in an effective field theory description \cite{Gubitosi:2012hu}. This effective field theory of dark energy allows one to describe models of dark energy and modified gravity using an operator expansion breaking Lorentz invariance explicitly\cite{Gleyzes:2013ooa,Piazza:2013coa,Bloomfield:2012ff,Silvestri:2013ne,Frusciante:2013zop,Bloomfield:2013cyf,Piazza:2013pua,Perenon:2015sla}\footnote{ Other methods for parameterising  dark energy/ modified gravity  have also been introduced in the last few years for models with second order equations of motion \cite{Baker:2012zs,Sawicki:2012re,Battye:2012eu,Battye:2015hza}. The comparison between the advantages of these different approaches is beyond the scope of the present paper.}. The coefficients of each operator are an unspecified function of time whose behaviour characterises each model at stake. Of course, this approach suffers from an enormous degeneracy, i.e. the appearance of many functions of time, with a multiplicity that grows
with the perturbative order. This means that the applicability and the comparison with observational results of this approach is restricted to low orders of perturbation theory
(typically once considers the background and linear levels). On the other hand, this
provides a very general framework.
For simple cases, such as the ones used in the Planck analysis \cite{Ade:2015rim}, one can hope to extract information about dark energy and modified gravity scenarios from the comparison with data.

The effective field theory approach also suffers from the fact that gravity acts on all scales, therefore implying that a perturbative approach at the cosmological level will usually not
be able to unravel properties
of dark energy or modified gravity models from the largest scales in the Universe to the Solar System ones. Such an extension is nevertheless crucial as one cannot be content with a large-scale description of dark energy or modified gravity which cannot be shown to be compatible with local measurements in the Solar system \cite{Bertotti:2003rm,Williams:2012nc}. The traditional approach to this issue, which involves building explicit models and screening mechanisms \cite{Khoury:2010xi}, complements the effective field theory description. For theories involving only
second-order equations of motion, the compatibility between dark energy or modified gravity with local tests relies on the known screening mechanisms: chameleon\cite{KhouryWeltman,Khoury:2003aq,Brax:2004qh} and Damour-Polyakov \cite{Damour:1994zq,Pietroni:2005pv,Hinterbichler:2010es,Olive:2007aj,Brax:2010gi}, K-mouflage \cite{Babichev:2009ee,Brax:2012jr} and Vainshtein \cite{Vainshtein:1972sx}.

In this paper, we focus on the K-mouflage mechanism \cite{Brax:2014wla,Brax:2014gra,Brax:2014yla,Barreira:2014gwa,Barreira:2015aea,Brax:2015lra} and examine its description in terms of the effective field theory of dark energy. K-mouflage models are $K$-essence models \cite{ArmendarizPicon:2000dh,ArmendarizPicon:2000ah} complemented with a universal coupling of the scalar field to matter. The coupling to matter changes the cosmological dynamics of the models compared to $K$-essence \cite{Brax:2014wla}. This can be exemplified by the looser link between the equation of state of dark energy $w_{\rm de}$ and the speed of scalar perturbations $c_s$ \cite{Brax:2015lra} with no imaginary $c_s$ when $w_{\rm de}<-1$.
Because we start from a specific scenario, the K-mouflage modified-gravity model,
which is fully defined at the nonlinear level, we obtain the exact nonlinear effective field
theory action and equations of motion.
It turns out that the effective field theory action up to quadratic order involves the first two derivatives of the K-mouflage function
$K(\tilde\chi)$, where $\tilde\chi$ is the reduced kinetic energy of the K-mouflage scalar whose presence affects gravity. In particular, the role of $K'(\tilde\chi)$, the first derivative of $K(\tilde\chi)$, is made explicit inasmuch as it appears explicitly in the modification of Newton's constant \cite{Brax:2014gra}
${\cal G}_{\rm N} = \bar{A}^2 (1+ \frac{2\beta^2}{\bar K'}) {\cal G}_{\rm N0}$ and needs to be large in screened situations. In the cosmological context, screening occurs in the early Universe when densities are large and $\bar K'\gg 1$. On the contrary, when dark energy and modified gravity effects matter at small redshift, the function $\bar K'$ tends to one and the K-mouflage field behaves like a massless scalar coupled to matter with a strength $\beta$. The coupling function $\bar A$ plays a crucial role in the cosmological evolution of Newton's constant and is constrained by the local time dependence of Newton's constant in the solar system providing a link between the cosmological and the local properties of K-mouflage.

The K-mouflage models, and therefore their effective action, can be reverse engineered to
accommodate  the screening requirement  $K'\gg 1$ in the early Universe
(to ensure that the dark energy is subdominant at early times) and
unscreening in sparse ones such as large cosmological scales in the recent past of the Universe.
We show how by specifying two functions $U(a)$ and $\bar A(a)$ at the background level,
we can reconstruct the whole K-mouflage theories
in the large-scale regime that applies to cosmological studies, down to cluster scales.
The function $U(a)$ is similar to the time variable of the background cosmology in the K-mouflage models expressed as a function of the scale factor $U(a) \sim t(a)$.
This new parameterization is very convenient as it allows us to normalize the models
today and to directly set the magnitude and the behaviour of the deviation from the
$\Lambda$-CDM cosmology through the function $\bar A(a)$.
In the case of K-mouflage scenarios, the cosmological
background and the growth of large-scale structures only probe the semi-axis
$\tilde\chi > 0$ whereas small subgalactic scales, such as the Solar System,
correspond to large negative $\tilde\chi$.
Therefore, these models are a simple example of theories where the
cosmological behaviour is decoupled from the small-scale behaviour\footnote{
One may be able to infer the negative-$\tilde\chi$ behaviour of $K(\tilde\chi)$ from
its positive-$\tilde\chi$ behaviour by requiring analyticity, but this is not very practical
as small deviations on the positive axis can be associated with large deviations on
the negative axis.}.

The dark energy effective action for models like K-mouflage does not carry new information which cannot be extracted from the original non-linear Lagrangian. In that respect,  the purpose of our paper is twofold. First we show that the complete K-mouflage theory can be defined not only by its Lagrangian but also by two functions of the cosmological scale factor. Given  these two functions, it is straightforward to reconstruct the original Lagrangian, to derive the cosmological background evolution and to deduce the effective dark energy effective theory at any order. This reconstruction can be easily used at second order in the effective theory, which is enough to analyse
all the linear observables of the models and eventually to compare them with data. Secondly,  the reconstruction mapping does not require any knowledge of the initial conditions in the early Universe and only requires the use
of the cosmological parameters like $\Omega_{\rm m0}$ now. This is a drastic advantage compared with integrating the dynamics from initial conditions as the parameters of the model such as $\cM$, the cut-off scale of the model, do not have to be adjusted by a trial and error procedure. This is particularly important when implementing the comparison with data in a systematic way using a modified version of CAMB for instance. The phenomenological work using the method expounded here is currently under way \cite{us}.

The paper is arranged as follows. In section~\ref{sec:K-mouflage}, we define the
K-mouflage models and recall the relevant equations of motion, at the background
and linear levels.
In section~\ref{sec:effective} we describe the effective field theory of K-mouflage.
In section~\ref{sec:Cosmological-pi}, we consider the cosmological dynamics at the
background and perturbative levels and we check that we recover the equations of
motion obtained from the usual K-mouflage action.
In section~\ref{sec:Parameterizing-K-mouflage},
we show how the whole cosmological regime of K-mouflage models can be reconstructed
from the knowledge of two functions, which provides a convenient parameterization
of such scenarios.
In section~\ref{sec:Numerical}, we present numerical results for models behaving like cubic K-mouflage.
We then conclude.
In the appendices we provide details about the scalar field description of the
quadratic effective action and about the causality of the theory.

\section{K-mouflage}
\label{sec:K-mouflage}

\subsection{Definition of the model}
\label{sec:def-K-mouflage}

We are interested in K-mouflage models, which we define by the action
\be
S = \int d^4 x \sqrt{-\tg} \left[ \frac{\tM_{\rm Pl}^2}{2} \tR + \cM^4 K(\tilde{\chi}) \right]
+ \int d^4 x \sqrt{-g} L_{\rm m}(\psi_i, g_{\mu\nu})
+ \int d^4 x \sqrt{-g} \frac{1}{4\alpha} F^{\mu\nu}F_{\mu\nu} ,
\label{S-def}
\ee
where $\tg_{\mu\nu}$ is the Einstein-frame metric and $g_{\mu\nu}$ the
Jordan-frame metric, given by the relation between Einstein- and Jordan-frame
line elements
\be
ds^2 = A^2(\varphi) d\ts^2 , \;\;\; g_{\mu\nu} = A^2(\varphi) \, \tg_{\mu\nu} ,
\label{conformal}
\ee
when we use the same coordinates in both frames, and
$A(\varphi)$ is the coupling function of the model.
Throughout this paper, we denote Einstein-frame quantities by a tilde to distinguish
them from Jordan-frame quantities.
The first and second parts in the action (\ref{S-def}) give the gravitational and scalar
field contributions, which are defined in the Einstein frame.
The K-mouflage Lagrangian of the scalar field $\varphi$ involves a non-canonical
kinetic term, defined by a dimensionless function $K(\tilde{\chi})$ with
\be
\tilde{\chi} = - \frac{\tg^{\mu\nu} \partial_{\mu}\varphi\partial_{\nu}\varphi}{2\cM^4} ,
\label{chi-def}
\ee
where $\cM^4$ is of the order of the dark energy scale today.
The third and fourth parts in the action (\ref{S-def}) give the matter and radiation
contributions, which are expressed in the Jordan frame,
and $\psi_i$ are various matter fields.

At the background level, we have the FLRW metrics
\be
d \bar{\ts}^2 = \bar{\tg}_{\mu\nu} dx^{\mu} dx^{\nu} = \tilde{a}^2(\tau)
[ - d \tau^2  + d{\bf x}^2 ]
\ee
and
\be
d \bar{s}^2 = \bar{g}_{\mu\nu} d x^{\mu} d x^{\nu}
= a^2(\tau) [- d\tau^2 + d{\bf x}^2 ] ,
\label{ds2-def}
\ee
where $\tau$ is the conformal time, and the scale factors are related by
\be
a = A(\bar\varphi) \tilde{a} .
\ee
Throughout this paper, we denote background quantities with a bar.

\subsection{Dynamics of K-mouflage cosmologies}
\label{sec:Dynamics-K-mouflage}

We briefly recall here the equations of motion that govern the dynamics of K-mouflage
cosmologies in the Jordan frame. We refer the reader to
\cite{Brax:2014wla,Brax:2014gra,Brax:2014yla,Brax:2015lra} for the derivations
from the action (\ref{S-def}).

\subsubsection{Cosmological background}
\label{sec:background-K-mouflage}

In the Jordan frame, the Friedmann equation takes the form
\be
3 M_{\rm Pl}^2 H^2 (1-\epsilon_2)^2 = \bar{\rho} + \bar{\rho}_{\rm rad}
+ \bar{\rho}_{\varphi}  \;\;\; \mbox{with} \;\;\;  M_{\rm Pl} = \bar{A}^{-2} \tM_{\rm Pl}^2 , \;\;\;
\epsilon_2 = \frac{d\ln\bar{A}}{d\ln a} ,
\label{Friedmann-2}
\ee
where we have introduced the scalar field energy density given by
\be
\bar{\rho}_{\varphi} = \bar{A}^{-4} \cM^4 ( - \bar{K} + 2 \bar{\tilde{\chi}} \bar{K}' )
\;\;\; \mbox{with} \;\;\; \bar{\tilde\chi} = \frac{\bar{A}^2}{2\cM^4}
\left( \frac{d\bar\varphi}{dt} \right)^2  .
\label{chi-dphi-dt}
\ee
Here $t$ is the Jordan-frame cosmic time, with $dt= a d\tau$, and $H=d\ln a/dt$ the
Jordan-frame Hubble expansion rate.
The matter and radiation energy densities obey the usual conservation laws, which give
\be
\bar\rho(a) = \frac{\bar\rho_0}{a^3} \;\;\; \mbox{and} \;\;\;
\bar\rho_{\rm rad}(a) = \frac{\bar\rho_{\rm rad 0}}{a^4} ,
\label{rho-matter-rho-rad}
\ee
and the Klein-Gordon equation satisfied by the scalar field writes as
\be
\frac{d}{dt} \left[ \bar{A}^{-2} a^3 \frac{d\bar\varphi}{dt} \bar{K}' \right] =
- a^3 \bar\rho \frac{d\ln\bar{A}}{d\bar\varphi} =
- \frac{\beta\bar{\rho}_0}{\tilde{M}_{\rm Pl}} ,
\label{KG-1}
\ee
where we defined
\be
\beta(a) \equiv \tilde{M}_{\rm Pl} \frac{d\ln \bar A}{d\bar\varphi} .
\label{def-beta-K}
\ee

\subsubsection{Linear perturbations}
\label{sec:linear-K-mouflage}

At the perturbative level, at linear order in the metric and density fluctuations, the Einstein
equations can be analysed in the Newtonian gauge,
\be
ds^2= a^2 \left[ - (1+2\Phi) d\tau^2 + (1-2\Psi) d{\bf x}^2 \right] ,
\label{ds2-Newtonian-gauge}
\ee
where the two Newtonian potentials $\Phi$ and $\Psi$ are not equal.
In the non-relativistic ($\Phi \ll 1$, $\Psi \ll 1$ and $v^2 \ll 1$) and small-scale
($k/aH \gg 1$) limits, the Einstein equations lead to
\be
\Phi= ( 1+\epsilon_1 ) \Psi_{\rm N} , \;\;\;
\Psi= ( 1-\epsilon_1 ) \Psi_{\rm N} , \;\;\; \mbox{with} \;\;\;
\epsilon_1 = \frac{2\beta^2}{\bar{K}'} , \;\;\;
\frac{\nabla^2}{a^2}\Psi_{\rm N} = \frac{\delta\rho}{2 M_{\rm Pl}^2} .
\label{Poisson-K}
\ee
In the last equation in (\ref{Poisson-K}), we used the fact that fluctuations of the scalar field
energy density are negligible as compared with fluctuations of the matter density \cite{Brax:2014yla,Brax:2015lra},
$\delta\rho_{\varphi} \ll \delta\rho$, in the small-scale limit. In a nutshell, this springs from the Klein-Gordon equation, which implies that
$\delta \varphi \sim (\beta a^2 / \bar K' \tilde M_{\rm Pl} k^2) \delta \rho$ when $k/a \gg H$.
This leads to $\delta \rho_\varphi \sim (\beta^2 a^2 H^2/\bar K' k^2) \delta\rho \ll \delta\rho$.

At the linear level, the effect of screening appears through the factor $\epsilon_1$ which measures the deviation of the two Newtonian potentials $\Phi$ and $\Psi$ from $\Psi_{\rm N}$. When $\epsilon_1\ll 1$, gravity is not modified and the two Newtonian potentials coincide. It turns out that $\epsilon_1\ne 0$ is also responsible for the change in the growth of structure for K-mouflage models. Indeed
the linear growing mode of  the density contrast, $D= \frac{\delta \rho}{\bar \rho}$, obeys the evolution equation
\be
\frac{d^2 D}{d(\ln a)^2} + \left( 2 + \frac{1}{H^2} \frac{d H}{d t} \right)
\frac{d D}{d\ln a} - \frac{3}{2} \Omega_{\rm m} (1+\epsilon_1) D = 0 ,
\label{D-linear-Jordan}
\ee
where the gravitational force is multiplied by the factor $(1+\epsilon_1)$ as
compared with the $\Lambda$-CDM cosmology, i.e the effective Newtonian constant is
\be
\frac{{\cal G}_{\rm N, eff}} {{\cal G}_{\rm N}}= 1+ \epsilon_1\label{G} ,
\ee
corresponding to a modification of gravity. At the linear level in perturbations, screening occurs when $\epsilon_1\ll 1$, which corresponds to backgrounds where $\bar{K}' \gg 1$, as for the early-Universe cosmological background.
It has to be noticed that the same factor $\epsilon_1$ appears
in the static case and (\ref{G}) is also valid. In this situation too, screening occurs where $\epsilon_1 \ll 1$, which turns out to be inside the K-mouflage radius where
GR is retrieved around a static object.

\section{Effective field theory}
\label{sec:effective}

We now describe how the K-mouflage model defined by the action (\ref{S-def})
can be recast in the effective field theory language. This can be useful to compare
theoretical predictions with observations of the background and linear regimes,
as data analysis may express such cosmological measurements in terms of the effective
field theory functions in order to remain as general as possible. Then, within each class
of scenario the parameters of the models must be determined to recover these few measured factors
(if such a solution exists).

One should be aware that the dark-energy effective-field-theory language which will be used throughout this paper is only a book-keeping device to analyse
cosmological perturbations around a given background cosmology. In the case of K-mouflage, the background cosmology is defined by the non-linear K-mouflage evolution equations whilst
the K-mouflage cosmological perturbations can be cast in the dark-energy effective-field-theory mould. As such the dark-energy effective field theory is only an expansion scheme
for the cosmological perturbations of a classical theory, which is either explicitly known as in the case of the K-mouflage models or parameterised in a model-independent approach by unknown time-dependent coupling functions for each operator appearing in the
effective Lagrangian. In the case of explicit models, the operators that appear are not necessarily all possible operators allowed by the generic cosmological symmetries (that are kept in a generic model-independent parameterisation) and we shall find that only one class of simple operators is generated by K-mouflage models.

The dark-energy effective field theory should be contrasted with the quantum effective field theory obtained by including radiative corrections
to the classical Lagrangian of a given model. The quantum effective field theory can be obtained by expanding in perturbation theory the classical Lagrangian around a given  background configuration and then calculating loop corrections to the classical action.   The quantum effective field theory of K-mouflage models will be analysed in a forthcoming publication. In a nutshell, the dimension-full scale which plays the role of an effective cut-off for K-mouflage models is ${\bar K'\ }^{1/4} \cM$. In this case, it can be shown that the classical K-mouflage Lagrangian is not renormalised and that for scales larger than the inverse of the effective cut-off, a small scale compared to astrophysical ones,  the finite correction terms to the classical action are all negligible. At these low energies, the particles of the standard model have all decoupled and it is legitimate to neglect the effect of the matter loops on the K-mouflage action. Hence it is perfectly legitimate to take as a starting point the classical and shift-symmetric K-mouflage action to develop  the classical dark-energy effective-field theory.

\subsection{Unitary gauge}
\label{sec:unitary}

In the general perturbed case, the scalar field $\varphi({\bf x},\tau)$ depends on both
time and position. However, we can make a change of coordinate to the
uniform-$\varphi$ slicing or unitary gauge, defining for instance the coordinates
${\bf x}_{(u)}$ and $\tau_{(u)}$ by
\be
{\bf x}_{(u)} \equiv {\bf x} , \;\;\; \tau_{(u)} \equiv \bar\varphi^{-1}[\varphi({\bf x},\tau) ] ,
\label{un}
\ee
where the subscript $(u)$ refers to the unitary gauge. Thus we assume that $\varphi$
can be used as a clock (which may not be valid on small nonlinear scales).
Then, by definition the scalar field $\varphi=\bar\varphi(\tau_{(u)})$ only depends on the
time $\tau_{(u)}$.
Thanks to the invariance under  diffeomorphisms of General Relativity,
the volume element transforms as $d^4 x \sqrt{-\tg} = d^4 x_{(u)} \sqrt{-\tg_{(u)}}$,
the Ricci scalar as $\tR=\tR_{(u)}$,  and the kinetic factor as
\be
\tilde{\chi} = - \chi_\ast(\tau_{(u)}) \, \tg^{00}_{(u)} \;\;\; \mbox{with} \;\;\;
\chi_\ast(\tau) = \frac{1}{2\cM^4} \left( \frac{d\bar\varphi}{d\tau} \right)^2 .
\ee
We also move to the unitary gauge in the Jordan frame of the matter action
(\ref{S-def}) which becomes
\be
S = \int d^4 x_{(u)} \sqrt{-\tg_{(u)}} \left[ \frac{\tM_{\rm Pl}^2}{2} \tR_{(u)}
+ \cM^4 K( - \chi_\ast \, \tg^{00}_{(u)} ) \right]
+ S_{\rm m}(\psi_i, g_{(u)\mu\nu}) +  S_{\rm rad}(F_{\mu\nu}, g_{(u)\mu\nu}) ,
\label{S-unit1}
\ee
with now
\be
g_{(u)\mu\nu} = \bar{A}^2 \, \tilde{g}_{(u)\mu\nu}  \;\;\; \mbox{and}
\;\;\; \bar{A} = A[\bar\varphi(\tau_{(u)})] .
\ee
Expanding in  the perturbation
$\delta\tilde{g}^{00}_{(u)} = \tilde{g}^{00}_{(u)} - \bar{\tilde{g}}^{00}_{(u)}$ of the metric,
this gives
\be
\begin{split}
S = & \int d^4 x_{(u)} \sqrt{-\tg_{(u)}} \left[ \frac{\tM_{\rm Pl}^2}{2} \tR_{(u)}
+ \cM^4 \sum_{n=0}^{\infty} \frac{\bar{K}^{(n)}(-\chi_\ast \bar{\tilde{g}}^{(00)}_{(u)})}
{n!} (-\chi_\ast \delta \tilde{g}^{00}_{(u)})^n \right] \\
& + S_{\rm m}(\psi_i, \bar{A}^2 \tg_{(u)\mu\nu})
+ S_{\rm rad}(F_{\mu\nu}, \bar{A}^2 \tg_{(u)\mu\nu})
\end{split}
\label{S-unit2}
\ee
where we have defined $\bar{K}^{(n)}=d^n K/d\tilde\chi^n (\bar{\tilde\chi})$ the derivative
of order $n$ of the kinetic function $K$ at the background value $\bar{\tilde\chi}$.
This action can be written under the general form of the Einstein-frame
effective-field-theory action of Gubitosi-Piazza-Vernizzi \cite{Gubitosi:2012hu}, as
\be
\begin{split}
S = & \int d^4 x \sqrt{-\tg_{(u)}} \left[ \frac{\tilde{M}_{\rm Pl}^2}{2} \tilde{R}_{(u)}
- \tilde{\Lambda}(\tau_{(u)}) - \tilde{c}(\tau_{(u)}) \tg^{00}_{(u)} + \sum_{n=2}^{\infty}
\frac{\tilde{M}_n^4(\tau_{(u)})}{n!} (\delta \tilde{g}^{00}_{(u)})^n \right] \\
& + S_{\rm m}(\psi_i, \tg_{(u)\mu\nu}/f(\tau_{(u)}))
+ S_{\rm rad}(F_{\mu\nu}, \tg_{(u)\mu\nu}/f(\tau_{(u)})) ,
\end{split}
\label{S-unit3}
\ee
with
\be
\begin{split}
& f(\tau_{(u)}) = \bar{A}^{-2} , \;\;\;
\tilde{\Lambda}(\tau_{(u)})  = - \cM^4 [ \bar{K}
+\chi_\ast \bar{K}' \bar{\tilde{g}}^{00}_{(u)} ] , \;\;\;
\tilde{c}(\tau_{(u)}) = \cM^4 \chi_\ast \bar{K}' , \\
& n \geq 2 : \;\; \tilde{M}_n^4(\tau_{(u)}) = \cM^4  (-\chi_\ast)^n \bar{K}^{(n)} .
\end{split}
\label{f-Lambda-c-Mn-Einstein}
\ee
When the model is a simple canonical scalar field with a cosmological constant,
\be
K(\tilde\chi) = -1 + \tilde\chi ,
\ee
we find that
\be
\tilde{\Lambda} = \cM^4, \;\;\; \tilde{c} = \cM^4 \chi_\ast , \;\;\;
n \geq 2 : \; \tilde{M}_n^4 = 0 ,
\ee
corresponding to both the cosmological constant and the kinetic energy of the scalar
field. The inclusion of higher order terms in $\tilde\chi$ in the K-mouflage Lagrangian
will imply that both $\tilde{\Lambda}$ and $\tilde{c}$ will take more complex forms,
with $\tilde{\Lambda}$ becoming time dependent.

\subsection{Jordan frame}
\label{sec:Jordan}

We can also write the action (\ref{S-unit1}) in the Jordan frame, using the
transformation laws
\be
\tilde{g}_{\mu\nu} = f g_{\mu\nu} , \;\;\;
\sqrt{-\tilde{g}} = f^2 \sqrt{-g} , \;\;\;
\tilde{R} = f^{-1} \left[ R - 3 \Box \ln f
- \frac{3}{2} g^{\mu\nu} \partial_{\mu}\ln f \partial_{\nu}\ln f \right] .
\ee
In the case of the unitary gauge, this simplifies because $f=f(\tau_{(u)})$ and
the action (\ref{S-unit1}) writes as
\be
\begin{split}
S = & \int d^4 x_{(u)} \sqrt{-g_{(u)}} \left[ \frac{f \tM_{\rm Pl}^2}{2} R_{(u)}
+ \frac{3}{4} f \tM_{\rm Pl}^2 g^{00}_{(u)} \left( \frac{d\ln f}{d\tau} \right)^2
+ f^2 \cM^4 K( - \chi_\ast f^{-1} g^{00}_{(u)} ) \right]  \\
& + S_{\rm m}(\psi_i, g_{(u)\mu\nu}) + S_{\rm rad}(F_{\mu\nu}, g_{(u)\mu\nu}) .
\end{split}
\label{S-J-unit1}
\ee
Expanding in  $\delta g^{00}_{(u)}$, as in Eq.(\ref{S-unit2}), we obtain an expression
similar to Eq.(\ref{S-unit3}),
\be
\begin{split}
S = & \int d^4 x \sqrt{-g_{(u)}} \left[ \frac{M_{\rm Pl}^2}{2} R_{(u)}
- \Lambda(\tau_{(u)}) - c(\tau_{(u)}) g^{00}_{(u)} + \sum_{n=2}^{\infty}
\frac{M_n^4(\tau_{(u)})}{n!} (\delta g^{00}_{(u)})^n \right] \\
& + S_{\rm m}(\psi_i, g_{(u)\mu\nu}) + S_{\rm rad}(F_{\mu\nu}, g_{(u)\mu\nu}) ,
\end{split}
\label{S-J-unit2}
\ee
with
\be
M_{\rm Pl}^2 = f \tilde{M}_{\rm Pl}^2 , \;\;\;
\Lambda = f^2 \tilde{\Lambda} , \;\;\;
c = f \tilde{c} - \frac{3}{4} M_{\rm Pl}^2 \left( \frac{d\ln f}{d\tau} \right)^2 , \;\;\;
M_n^4 = f^{2-n} \tilde{M}_n^4 .
\label{Lambda-c-J-def}
\ee
The expression (\ref{S-J-unit2}) gives the Jordan-frame effective-field-theory
action that corresponds to K-mouflage models, in the unitary gauge.
In the usual effective field theory approach, the action is only defined as a perturbative
expansion up to some low order, such as the quadratic level $n=2$ if we consider
the equations of motion up to linear order in the field fluctuations.
This provides a general framework, as at each order one considers all possible
operators, that applies to many different scenarios. On the other hand, the
number of operators grows with the order and the theory is not specified beyond the
truncation level.

In this paper, since we focus on the K-mouflage models defined by the fully
nonlinear explicit action (\ref{S-def}), we have an explicit closed form (\ref{S-J-unit1})
for the effective action and we obtain the expansion (\ref{S-J-unit2}) at all orders.
Moreover, it happens that the K-mouflage action (\ref{S-def}) is simple enough
to  generate only a small subset of all possible operators at high orders.
In fact, at each order we only have the single operator $(\delta g^{00})^n$.
This is a key signature of K-mouflage scenarios amongst the class of theories coupled to matter  in the context
of dark energy effective field theories. If observations showed that some other operators cannot be set
to zero, this would rule out all K-mouflage models.

\subsection{Equivalence with the scalar field description}
\label{sec:scalar-field-pi}

Although not transparent, the dynamics of the models defined by an effective action
such as (\ref{S-J-unit2}) are equivalent to a modification of General Relativity (GR)
by a scalar degree of freedom.
The scalar field which appears in the original definition of K-mouflage models can be
unravelled by performing the changes \cite{Cheung:2008}
\be
\tau_{(u)} \rightarrow \tau + \pi ({\bf x},\tau) , \;\;\; g^{00} _{(u)} \rightarrow \partial_{\mu}(\tau+\pi)
\partial_{\nu}(\tau+\pi) g^{\mu\nu} ,
\label{Stueckelberg}
\ee
in the effective action that we have just obtained. We have the non-linear transformation
$\varphi({\bf x},\tau)= \bar \varphi (\tau+\pi ({\bf x},\tau))$ and the
perturbations of the scalar field are captured by $\pi$ in a non-linear way.
At the linear level we have
\be
\delta\varphi = \frac{d\bar\varphi}{d\tau} \pi .
\label{linear-phi-pi}
\ee
The metric $g_{\mu\nu}$ is the metric associated to the coordinates $({\bf x},\tau)$,
where $\tau$ is the background conformal time, and $\pi=0$ for the cosmological
background.
This ``Stueckelberg trick'' (\ref{Stueckelberg}) restores the diffeomorphism invariance
of the action, as an infinitesimal change of time can be absorbed through $\pi$
as $\tau \rightarrow \tau + \xi^0$, $\pi \rightarrow \pi-\xi^0$.

In the appendix \ref{app:S2-def} we consider this change at the level of the
quadratic effective action, as in the usual effective field theory approach.
Here, because we actually know the action up to all orders since we start from the
explicit nonlinear action (\ref{S-def}) associated with K-mouflage models,
we start from the full K-mouflage action (\ref{S-J-unit1}) in the unitary gauge and
change the time slicing.
Both approaches are equivalent and lead to the same dynamical equations at the linear
order of perturbation theory.

After the transformation (\ref{Stueckelberg}), the Jordan frame action (\ref{S-J-unit1}), which was written in the unitary gauge where the time $\tau_{(u)}$ was specified by
(\ref{un}) explicitly, becomes
\be
\begin{split}
S = & \int d^4 x \sqrt{-g} \biggl \lbrace \frac{f(\tau + \pi)\tM_{\rm Pl}^2}{2} R
+ \frac{3}{4} f(\tau + \pi) \tM_{\rm Pl}^2 g^{\mu\nu} \partial_\mu (\tau+\pi)
\partial_\nu(\tau+\pi) \left[ \frac{d\ln f}{d\tau}(\tau+\pi) \right]^2 \\
&  + f^2(\tau+\pi) \cM^4 K \left[ f^{-1}(\tau+\pi) \chi_\star(\tau+\pi) g^{\mu\nu}
\partial_\mu (\tau+\pi) \partial_\nu(\tau+\pi) \right] \biggl \rbrace + S_{\rm m} + S_{\rm rad} .
\end{split}
\label{S-pi-J}
\ee
This gives the full nonlinear K-mouflage action in terms of the scalar field $\pi$ in the general coordinate system $({\bf x},\tau)$. In the appendix~\ref{app:S2-def}, we
analyse the action at second order in $\pi$, using the conformal Newtonian gauge for
the metric $g_{\mu\nu}$.

Because the scalar field $\pi$ has been re-introduced through the
unitary gauge effective action (\ref{S-J-unit1}) and the Stueckelberg trick
(\ref{Stueckelberg}), which have a very general scope but are not specifically devised
for K-mouflage models, the action (\ref{S-pi-J}) is significantly more complex than the
original action (\ref{S-def}).
In particular, the scalar field $\pi$ is not identical to the original scalar field $\varphi$,
though at linear level it corresponds to $\delta\varphi$.

The explicit expression (\ref{S-pi-J}) suggests that it may be difficult to extend the
effective field theory analysis to high perturbative orders, unless one can perform
some field redefinitions within specific classes of models, as even simple models such
as K-mouflage give rise to many terms as one expand in $\pi$.
On the other hand,  effective field theories are mainly used to study the linear
regime or low orders of perturbation theory for which the number of terms is reasonable.

\section{Cosmological dynamics from the effective field theory actions}
\label{sec:Cosmological-pi}

We now describe how the equations of motions that determine the K-mouflage cosmology
at the background and linear levels, which we recalled in
section~\ref{sec:Dynamics-K-mouflage}, can be recovered from the effective field theory
actions.

\subsection{Background}
\label{sec:Background-pi}

We can directly obtain the cosmological background dynamics from the unitary gauge
action (\ref{S-J-unit2}), using the metric (\ref{ds2-def}) which also describes the unitary
gauge in the homogeneous case.
Because the Jordan-frame Planck mass $M_{\rm Pl}$ depends on time,
the functional derivative of the first term in Eq.(\ref{S-J-unit2}) yields
\be
\frac{1}{\sqrt{-g}} \frac{\delta(\sqrt{-g} f R)}{\delta g^{\mu\nu}} = f G_{\mu\nu}
- \nabla_{\mu}\nabla_{\nu} f + g_{\mu\nu} \Box f ,
\ee
and the Einstein equations read as
\be
\begin{split}
& \tilde{M}_{\rm Pl}^2 \left[ f G_{\mu\nu} - \delta^0_{\mu} \delta^0_{\nu}
\left( \ddot{f} - {\cal H} \dot{f} \right) + \delta^i_{\mu} \delta^i_{\nu} {\cal H} \dot{f}
- g_{\mu\nu} a^{-2} \left( \ddot{f} + 2 {\cal H} \dot{f} \right) \right]
+ \left(\Lambda - \frac{c}{a^2} \right) g_{\mu\nu} \\
& - 2 c \delta^0_{\mu} \delta^0_{\nu}
= T_{\mu\nu} + T_{{\rm rad}\mu\nu} ,
\end{split}
\label{EE-unitary}
\ee
where $T_{\mu\nu}$ and $T_{{\rm rad}\mu\nu}$  are the matter and radiation
energy-momentum tensors. Throughout this paper we denote with a dot derivatives
with respect to the conformal time $\tau$, $\dot{f} = \partial f/\partial \tau$ and
$\ddot{f} = \partial^2 f/\partial\tau^2$ [here we actually have $\dot{f} = df/d\tau$].

The (00) component gives the Friedmann equation
\be
3 M_{\rm Pl}^2 \left( {\cal H}^2 + {\cal H} \frac{d\ln f}{d\tau} \right) = a^2 \bar{\rho}
+ a^2 \bar{\rho}_{\rm rad} + a^2 \Lambda + c .
\label{Friedmann-conformal-1}
\ee
Changing time coordinate from the conformal time $\tau$ to the Jordan-frame cosmic
time $t$, defined as $dt = a d\tau$ (which is different from the Einstein-frame time
$\tilde{t}$ given by $d\tilde{t} = \tilde{a} d\tau=\bar{A}^{-1} d t$), this reads as
\be
3 M_{\rm Pl}^2 \left( H^2 + H \frac{d\ln f}{dt} \right) = \bar{\rho} + \bar{\rho}_{\rm rad}
+ \Lambda + \frac{c}{a^2} .
\label{Friedmann-1}
\ee
Then, substituting the expressions of $\Lambda$ and $c$ obtained in
(\ref{Lambda-c-J-def}), we recover the Friedmann equation (\ref{Friedmann-2})
recalled in section~\ref{sec:background-K-mouflage}, which was
derived in \cite{Brax:2015lra} in the Jordan frame from the original action (\ref{S-def}).

\subsection{Perturbations}
\label{sec:Perturbations-pi}

To go beyond the background dynamics, we consider the fully nonlinear
effective-field-theory action (\ref{S-pi-J}) expressed in terms of the scalar field
$\pi$.
This allows us to derive the Einstein equations and the Klein-Gordon equation for $\pi$.
The Einstein equations read as
\be
\begin{split}
& \tilde M^2_{\rm Pl} [ f G_{\mu\nu} - \nabla_{\mu}\nabla_{\nu} f + g_{\mu\nu} \Box f ]
- g_{\mu\nu} \left[ \frac{3}{4} f \tM_{\rm Pl}^2 g^{\rho\sigma} \partial_\rho \tau_\pi
\partial_\sigma\tau_\pi \left( \frac{\dot{f}}{f}\right)^2 + f^2 \cM^4 K \right] \\
& + \frac{3}{2} f \tM_{\rm Pl}^2 \partial_\mu\tau_\pi \partial_\nu\tau_\pi
\left( \frac{\dot{f}}{f} \right)^2 - 2 f^2 \cM^4 K' a^2 \bar{\tilde\chi} \partial_\mu \tau_\pi
\partial_nu \tau_\pi =T_{\mu\nu} + T_{\rm {rad}\mu\nu} ,
\end{split}
\label{EE-pi-nl}
\ee
where we have defined the shifted time coordinate $\tau_\pi$ and the kinetic
factor $X$,
\be
\tau_\pi =\tau + \pi , \;\;\; X = - a^2 \bar{\tilde\chi} g^{\rho\sigma} \partial_\rho \tau_\pi
\partial_\sigma\tau_\pi ,
\label{tau-pi-X-def}
\ee
and we use $f^{-1} \chi_\star = a^2 \bar{\tilde\chi}$.
In Eqs.(\ref{EE-pi-nl}) and (\ref{tau-pi-X-def}), all explicit functions of time,
$f$, $\dot{f}$, $a^2$ and $\bar{\tilde\chi}$, are evaluated at the shifted time
$\tau_\pi$, which involves the scalar field $\pi$.
The kinetic functions $K$ and $K'$ are evaluated at the point $X$, which also involves the
scalar field $\pi$.

The Klein Gordon equation for $\pi$ can be obtained similarly,
\be
\begin{split}
& \tilde M^2_{\rm Pl} \dot{f} \left[ \frac{R}{2} +\frac{3}{4} g^{\mu\nu}\partial_\mu \tau_\pi
\partial_\nu \tau_\pi \left( \frac{\dot{f}}{f} \right)^2 + \frac{3}{2} g^{\mu\nu}\partial_\mu
\tau_\pi \partial_\nu \tau_\pi \left( \frac{\ddot{f}}{f} - \frac{\dot{f}^2}{f^2} \right) \right]
+ 2 f \dot{f} {\cal M}^4 K - f^2 {\cal M}^4 K' a^2 \\
& \times \left( \dot{\bar{\tilde\chi}} + 2{\cal H}\bar{\tilde\chi} \right)
g^{\mu\nu}\partial_\mu \tau_\pi \partial_\nu \tau_\pi
= 2 \nabla_\mu \left[ \frac{3}{4} f \tilde{M}_{\rm Pl}^2 g^{\mu\nu} \partial_\nu \tau_\pi
\left( \frac{\dot{f}}{f} \right)^2 - f^2 {\cal M}^4 K' a^2 \bar{\tilde\chi} g^{\mu\nu}
\partial_\nu \tau_\pi \right] ,
\end{split}
\label{KG-pi-nl}
\ee
where again the functions of time, $f$, $\dot{f}$, $a^2$, $\bar{\tilde\chi}$ and
$\dot{\bar{\tilde\chi}}$, are evaluated at the shifted time $\tau_\pi$, while the
kinetic functions $K$ and $K'$ are evaluated at the point $X$.

At the background level, the Einstein equations are obtained by putting $\pi\equiv 0$
and we retrieve Eq.(\ref{Friedmann-2}).
The Klein-Gordon equation reads when $\pi\equiv 0$
\be
\begin{split}
& 3\tilde M^2_{\rm Pl} \dot{f} \left[ {\cal H}^2 + \dot{\cal H} - \frac{\dot{f}^2}{4f^2}
+ \frac{\ddot{f}}{2 f} + {\cal H} \frac{\dot{f}}{f} \right] + 2 f \dot{f} a^2 {\cal M}^4 \bar{K} \\
& - a^2 f^2 {\cal M}^4 \bar{K}' \left( \dot{\bar{\tilde{\chi}}} + 6 {\cal H} \bar{\tilde\chi}
+ 4 \frac{\dot{f}}{f} \bar{\tilde\chi} \right)
- 2 a^2 f^2 {\cal M}^4 \bar{K}'' \bar{\tilde\chi} \dot{\bar{\tilde\chi}} = 0 ,
\end{split}
\label{KG-pi-0-background}
\ee
which can be proved to be  identically satisfied by using the Klein-Gordon and the
Friedmann equations (\ref{KG-1}) and (\ref{Friedmann-2}) of K-mouflage.
Thus, the Klein-Gordon equation (\ref{KG-1}) of the K-mouflage field $\varphi$ in the
Jordan frame and conformal time reduces to
\be
f^2 {\cal M}^4 \bar{K}' \left( 3 \frac{\dot{f}}{f} \bar{\tilde\chi} + 6 {\cal H}  \bar{\tilde\chi}
+ \dot{\bar{\tilde\chi}} \right) + 2  f^2 {\cal M}^4 \bar{K}'' \bar{\tilde\chi}
\dot{\bar{\tilde\chi}} = \frac{\dot{f}}{2 f} \bar{\rho} ,
\ee
where we substituted $\bar{A}=f^{-1/2}$.
The Friedman equation and its time derivative are given by
\be
3 \tilde{M}^2_{\rm Pl} f {\cal H}^2 = a^2 (\bar\rho + \bar\rho_{\rm rad} + \bar\rho_{\varphi} )
- 3 \tilde{M}^2 _{\rm Pl} f \left( {\cal H} \frac{\dot{f}}{f} + \frac{\dot{f}^2}{4 f^2} \right)
\ee
and
\be
-2 \tilde{M}^2_{\rm Pl} f \dot{\cal H} = a^2 \left( \frac{\bar\rho + \bar\rho_{\rm rad}
+ \bar\rho_{\varphi}}{3} + \bar{p}_{\rm rad} + \bar{p}_\varphi \right)
+ 2 \tilde{M}_{\rm Pl}^2 f \left( -\frac{\dot{f}^2}{2f^2} + \frac{\ddot{f}}{2f} \right) ,
\ee
where the scalar energy density and pressure are given by
\be
\bar\rho_\varphi = f^2{\cal M}^4 ( 2 \bar{\tilde\chi} \bar{K}' - \bar{K} ) , \;\;
\bar{p}_\varphi = f^2 {\cal M}^4 \bar{K} .
\ee
These identities combine to yield the Klein-Gordon equation (\ref{KG-pi-0-background})
for $\pi$ at the background level.

At the perturbative level, to linear order in the equations of motion, the Einstein equations
can be analysed in the Newtonian gauge (\ref{ds2-Newtonian-gauge}).
The time-time component of Einstein's equations (\ref{EE-pi-nl})
can be obtained  by noting that
\be
\bar G_{00}= 3 {\cal H}^2, \;\;\;
\delta G_{00}= -6 {\cal H} \partial_\tau \Psi +2 \nabla^2 \Psi ,
\ee
where $\nabla^2= \nabla_i \nabla^i$ is the flat-space Laplacian.
In the sub-horizon limit of perturbation theory, the gradient terms dominate and we obtain
in the quasi-static approximation
\be
\frac{\nabla^2}{a^2} \left( \Psi - \frac{\dot{f}}{2 f} \pi \right) =
\frac{\delta\rho}{2 M_{\rm Pl}^2}  ,
\ee
which is nothing but the Poisson equation relating the Newtonian potential $\Psi_{\rm N}$
to the matter perturbations (here we used that $\delta\rho_{\varphi} \ll \delta\rho$ in the
small-scale limit). The spatial part of Einstein's equations (\ref{EE-pi-nl}), using
\be
i \neq j : \;\;\; \delta G_{ij}= \partial_i \partial_j (\Psi -\Phi) ,
\ee
gives
\be
\partial_i\partial_j \left(\Psi -\Phi - \frac{\dot{f}}{f} \pi \right) = 0 ,
\ee
in the non-relativistic limit $v^2 \ll 1$, where $v$ is the velocity of the matter fluid.
At the linear order of perturbation theory and in the non-relativistic approximation
we have therefore
\be
\Phi = \Psi_{\rm N} - \frac{\dot{f}}{2f} \pi , \;\;\;\;\;
\Psi = \Psi_{\rm N} + \frac{\dot{f}}{2f} \pi ,
\label{Poisson-Phi-Psi-pi}
\ee
where $\Psi_{\rm N}$ is the Newtonian potential given by the last expression in
Eq.(\ref{Poisson-K}).

Similarly, the linearised Klein-Gordon equation in $\pi$ at first order can be deduced
from Eq.(\ref{KG-pi-nl}) by using
\be
a^2 \delta R= 4 \nabla^2 \Psi - 2 \nabla^2 \Phi
\ee
in the quasi-static and sub-horizon limit, leading to the linear relationship
\be
\frac{\dot{f}}{f} \pi = -\frac{4 \beta^2}{\bar{K}'} \Psi_{\rm N} ,
\ee
where $\beta$ was defined in Eq.(\ref{def-beta-K}).
This gives back Eq.(\ref{Poisson-K}), with a relative modification of the gravitationall
potentials from Newtonian gravity given by the factor $\epsilon_1=2\beta^2/\bar{K}'$.
In particular, we retrieve that screening occurs when $K' \gg 1$.
These results coincide with the ones in \cite{Brax:2015lra}. A comparison with the derivation from
the second-order effective action is given in the appendix.
This also leads to the relation between the effective field theory field $\pi$ and
the original scalar field $\varphi$ as
\be
\pi = - \frac{2f}{\dot{f}} \frac{\beta}{\tilde{M}_{\rm Pl}} \delta\varphi ,
\label{pi-delta-phi}
\ee
where $\delta\varphi = \varphi - \bar\varphi$ is the fluctuation of the scalar field
$\varphi$.
Equation (\ref{pi-delta-phi}), which agrees with Eq.(\ref{linear-phi-pi}),
 only holds at the linear level in the field fluctuations.

\section{Parameterizing K-mouflage}
\label{sec:Parameterizing-K-mouflage}

The K-mouflage models are usually defined by the kinetic function $K(\tilde\chi)$
and the coupling function $A(\varphi)$ that enter the K-mouflage action
(\ref{S-def}). However, we may instead look for a parameterization of these K-mouflage
models by two functions of the scale factor $a$, which would provide an implicit
definition of $K(\tilde\chi)$ and $A(\varphi)$.
This is similar to the tomographic approach developed in \cite{Brax:2011aw} for chameleon
models where, instead of the function $f(R)$ in $f(R)$-theories or the scalar field
potential $V(\varphi)$ and coupling $A(\varphi)$ in Dilaton and Symmetron models,
one defines the model through $m^2(a)$ and $\beta(a)$ [where
$m^2=\partial^2 V_{\rm eff}/\partial\varphi^2$ and
$\beta/\tilde{M}_{\rm Pl}= d\ln A/d\varphi$],
which allows one to reconstruct $V(\varphi)$ and $A(\varphi)$.

\subsection{Properties of the K-mouflage fields and characteristic functions}
\label{sec:Properties-of-K-mouflage}

Before devising such an implicit definition, it is useful to recall a few properties satisfied
by the kinetic and coupling functions.
We refer the reader to \cite{Brax:2014wla,Brax:2014yla,Brax:2014gra,Barreira:2015aea}
for the derivation of these constraints.
Over all the cosmological evolution we require
\be
\bar{A}>0 , \;\;\; \bar{K}' >0, \;\;\; \frac{d\bar\varphi}{dt} < 0 , \;\;\;
\beta > 0 , \;\;\; \epsilon_2 < 0 .
\label{signs-convention}
\ee
The condition $\bar{A}>0$ corresponds to the constraint that $A$ does not vanish,
so that the conformal transformation (\ref{conformal}) is always well defined. The constraint $\bar{K}'>0$ ensures that there are no
ghosts.
The scalar-field Lagrangian in the action (\ref{S-def}) being even in $\varphi$,
the sign of $\varphi$ can be absorbed in the sign of $\beta$ and we choose the
convention $\beta>0$. The coupling strength $\beta$ does not vanish and keeps
a positive sign if the scalar field is a monotonic function of time. In practice, the field
excursion in units of the Planck mass is modest, from the primordial universe
$a \rightarrow 0$ until now $a=1$, so that $\beta$ does not vary much.
Then, the combination $\bar{K}'>0$ and $\beta>0$ implies $d\bar\varphi/dt <0$
(as can be seen from the Klein-Gordon equation) and $\epsilon_2<0$.

In addition, we have the high-redshift behaviours
\be
a \rightarrow 0 :  \;\;\; \bar{A} \rightarrow A_\ast , \;\;\; \epsilon_2 \rightarrow 0 ,
\;\;\; \bar{K}' \rightarrow +\infty ,
\;\;\; \bar{\tilde\chi} \rightarrow +\infty , \;\;\; \bar\varphi \rightarrow \varphi_\ast .
\label{a-0-A-Kp}
\ee
The finite limit of $\bar{A}$ means that the Einstein and Jordan frames become
equivalent at early time so that we can recover the standard cosmology, without
a fifth force.
This also implies that $\epsilon_2$ goes to zero.
The condition $\bar{K}' \rightarrow +\infty$ ensures that the dark energy density
becomes subdominant with respect to the matter density at early time.
The marginal case where $\bar{K}'$ goes to a constant $K_\ast \gg 1$,
so that in the early matter era the dark energy density is a small fixed fraction of
the matter density, is also possible but we do not explicitly consider this case here
(it can be obtained as a limit of the range of models we consider).
The condition $\bar{\tilde\chi} \rightarrow +\infty$ is a natural consequence of
$\bar{K}' \rightarrow +\infty$ if we consider kinetic functions that have no singularity
at a finite value of $\tilde\chi$.
Note that because we look for a tomographic reconstruction of the K-mouflage model,
we restrict ourselves to models where $\tilde\chi(a)$ is a monotonic decreasing function
of $a$ to avoid reconstructing ill-defined multi-valued functions.
Although the time derivative of the scalar field, $d\bar\varphi/dt$, diverges at
early time like $\sqrt{\bar{\tilde\chi}}$, the scalar field goes to a finite value.
This is again due to the requirement that we recover the standard cosmology
at early time, which implies that the scalar kinetic energy cannot grow too fast
at high redshift.

At late time in the future, far in the dark energy era, the scalar field
$\bar\varphi$ will again converge to a finite value and we have
\be
a \rightarrow \infty: \;\;\; \bar{\tilde\chi} \rightarrow 0 , \;\;\;
\bar\varphi \rightarrow \varphi_{\infty} , \;\;\;
\bar{A} \rightarrow A_{\infty} , \;\;\; \epsilon_2 \rightarrow 0 .
\label{chi-phi-A-infty}
\ee
Finally, the kinetic function must satisfy the property
\be
\begin{split}
& \sqrt{\tilde\chi} K'(\tilde\chi) \;\;\;
\mbox{is a monotonic increasing function of $\tilde\chi$ over}
\;\; \tilde\chi > 0 , \;\;\; K'+2\tilde\chi K'' >0 ,  \\
& \mbox{and} \;\;\; \sqrt{\tilde\chi} K'(\tilde\chi) \rightarrow +\infty \;\;\;
\mbox{for} \;\;\; \tilde\chi \rightarrow +\infty .
\end{split}
\label{W+-cond}
\ee
This ensures that the system is well defined up to arbitrarily high redshifts,
when $\bar\rho\rightarrow+\infty$, and that there are neither small-scale instabilities
nor causality problems.

In this paper we focus on the cosmological background and large-scale perturbations,
which correspond to $\tilde\chi>0$. The kinetic function must also satisfy similar
properties on the semi-axis $\tilde\chi<0$, which corresponds to the small-scale
quasi-static regime, but we do not need to consider this here.

\subsection{Building a simple parameterization in terms of the scale factor}
\label{sec:Building-parameterization}

\subsubsection{The parameterisation}

In order to build an implicit parameterization we must consider the equations of motion satisfied
by the system at the background level, namely the Friedmann equation
(\ref{Friedmann-2}), the usual conservation equations for the matter and radiation
components (\ref{rho-matter-rho-rad}), and the Klein-Gordon equation (\ref{KG-1})
satisfied by the scalar field.
From the second relation (\ref{chi-dphi-dt}) and the sign convention
(\ref{signs-convention}) we have
\be
\frac{d\bar\varphi}{dt} = - \frac{\sqrt{2\bar{\tilde\chi}\cM^4}}{\bar{A}} , \;\;\;
\frac{d\bar\varphi}{d\ln a} = - \frac{\sqrt{2\bar{\tilde\chi}\cM^4}}{\bar{A}H} , \;\;\;
\frac{\beta}{\tilde{M}_{\rm Pl}} =
- \frac{\epsilon_2\bar{A}H}{\sqrt{2\bar{\tilde\chi}\cM^4}} ,
\label{dphi-dt-dphi-dlna}
\ee
and the Klein-Gordon equation (\ref{KG-1}) reads
\be
\frac{d}{d\ln a} \left[ \bar{A}^{-3} a^3 \sqrt{2\bar{\tilde\chi}\cM^4} \bar{K}' \right]
= - \frac{\epsilon_2 \bar{A}\bar\rho_0}{\sqrt{2\bar{\tilde\chi}\cM^4}} .
\label{KG-2}
\ee
This equation suggests that a simple parameterization of the K-mouflage model
is provided by the two functions of the scale factor
\be
U(a) \equiv a^3 \sqrt{\bar{\tilde\chi}} \bar{K}' \;\;\; \mbox{and} \;\;\; \bar{A}(a) ,  \;\;\;
\mbox{with} \;\;\; \bar{A}_0 \equiv \bar{A}(a=1) = 1 ,
\label{U-A-def}
\ee
where we choose to normalize the conformal factor $\bar{A}$ to unity today.
Given $\bar{A}(a)$ we obtain at once $\epsilon_2(a)$ from its definition
(\ref{Friedmann-2}) and the Klein-Gordon equation (\ref{KG-2}) yields
\be
\sqrt{\bar{\tilde\chi}} = - \frac{\bar\rho_0}{\cM^4} \,
\frac{\epsilon_2 \bar{A}^4}{2U (-3\epsilon_2+\frac{d\ln U}{d\ln a} )}
\label{sqrt-chi-a}
\ee
{\bf This gives an explicit parameterization of $\bar{\tilde \chi}(a)$.}

\subsubsection{Normalising $\cM$}

The full reconstruction of the Lagrangian $K(\chi)$ and the coupling function $A(\varphi)$ can only be achieved when the scale $\cM$ has been specified in terms of the present-time
cosmological parameters.
This can be achieved using  the Friedmann equation (\ref{Friedmann-2}) which writes as
\be
\frac{H^2}{H_0^2} = \frac{\bar{A}^2}{(1-\epsilon_2)^2}
 \left[ \frac{\Omega_{\rm m0}}{a^3} +  \frac{\Omega_{\rm rad 0}}{a^4}
+ \Omega_{\varphi 0} \bar{A}^{-4} \frac{\bar{K} - 2 \bar{\tilde\chi} \bar{K}'}
{\bar{K}_0 - 2\bar{\tilde\chi}_0\bar{K}'_0} \right] ,
\label{Friedmann-H-1}
\ee
where we defined the cosmological density parameters as
$\Omega_i(a)=\bar\rho_i(a)/\rho_{\rm crit}(a)$ and
$\rho_{\rm crit}(a)=3 M_{\rm Pl}^2(a) H^2(a)$.
In particular, from the scalar field energy density (\ref{chi-dphi-dt}) we have
\be
\Omega_{\varphi 0} = \frac{\cM^4}{\bar\rho_0} \, \Omega_{\rm m0} \,
(-\bar{K}_0+2\bar{\tilde\chi}_0\bar{K}'_0) .
\label{Omega-phi0-M4}
\ee

To close the system we need to specify the scale $\cM^4$ in terms of the
cosmological parameters today, as we wish to normalize the K-mouflage model
at $z=0$. To make the Friedmann equation (\ref{Friedmann-2}) look closer to the
$\Lambda$-CDM cosmology, we introduce an effective dark energy density defined by
\be
3 \tilde{M}_{\rm Pl}^2 \bar{A}^{-2} H^2 = \bar{\rho} + \bar{\rho}_{\rm rad}
+ \bar{\rho}_{\rm de} .
\ee
(However we keep the explicit time dependence of the Planck mass on the
left hand side as the Planck scale is intrinsically time-dependent in the Jordan frame and this effect could be detected by Solar System experiments.)
 This yields explicitly
\be
\bar{\rho}_{\rm de} \equiv \bar{\rho}_{\varphi} +
\frac{2\epsilon_2 - \epsilon_2^2}{(1-\epsilon_2)^2}
\left( \bar{\rho} + \bar{\rho}_{\rm rad} + \bar{\rho}_{\varphi} \right) .
\label{rho-de-Jordan}
\ee
The Jordan-frame dark-energy density evolves as
\be
\frac{d\bar{\rho}_{\rm de}}{d t} = - 3 H \left( \bar{\rho}_{\rm de}
+ \bar{p}_{\rm de} \right ) ,
\label{conserv-de-Jordan}
\ee
where we have defined the Jordan-frame dark-energy pressure as
\be
\bar{p}_{\rm de} = \bar{p}_{\varphi} +
\frac{\epsilon_2}{1-\epsilon_2} ( \bar{p}_{\rm rad} + \bar{p}_{\varphi} )
+ \left( \epsilon_2 - \frac{2}{1-\epsilon_2} \frac{d\epsilon_2}{d\ln a} \right)
\frac{\bar{\rho} + \bar{\rho}_{\rm rad} + \bar{\rho}_{\varphi}}{3 (1-\epsilon_2)^2} .
\label{p-de-Jordan}
\ee
Notice that the coupling to matter modifies the usual expressions for the dark-energy
pressure and energy density from the $K$-essence models. Similarly,
the effective dark-energy equation of state in the Jordan frame is simply defined as
\be
w_{\rm de} = \bar{p}_{\rm de} / \bar{\rho}_{\rm de} .
\label{w-Jordan-def}
\ee
This differs significantly from the $K$-essence equation of state
$w_{\varphi}$ defined
as
\be
w_{\varphi} = \frac{\bar{p}_{\varphi}}{\bar{\rho}_{\varphi}} = \frac{\bar{K}}{2\bar{\tilde\chi}
\bar{K}'-\bar{K}}.
\label{wphi-def}
\ee
In particular, the speed of scalar perturbations $c_s^2$ in K-mouflage models is not directly
related to $(1+ w_{\rm de})$ like for $K$-essence models \cite{Vikman:2004dc}.
We will see that models can cross the phantom divide $w_{\rm de}<-1$ in the past of the
Universe without any instability, as $ c_s^2$ remains positive.

We can use the Friedmann equation to normalize
\be
\Omega_{\rm m} + \Omega_{\rm rad} + \Omega_{\rm de} =
\frac{\Omega_{\rm m} + \Omega_{\rm rad} + \Omega_{\varphi}}{(1-\epsilon_2)^2} ,
\ee
whence
\be
\Omega_{\varphi 0} = (1-\epsilon_{2,0})^2 \Omega_{\rm de 0}
- (2\epsilon_{2,0}-\epsilon_{2,0}^2) (\Omega_{\rm m0}+\Omega_{\rm rad0})
\;\;\; \mbox{with} \;\;\; \epsilon_{2,0} \equiv \epsilon_2(a=1) .
\label{Omega-phi-0-def}
\ee
This provides the value of the cosmological parameter $\Omega_{\varphi 0}$ today
in terms of the standard cosmological parameters
$\{\Omega_{\rm m0},\Omega_{\rm rad 0},\Omega_{\rm de0} \}$ that would be
derived from the Friedmann equation today assuming a $\Lambda$-CDM cosmology.
In realistic models, which remain close to the $\Lambda$-CDM scenario
and must satisfy Solar System tests, we have $|\epsilon_2| \lesssim 0.01$
and $\Omega_{\varphi 0} \simeq \Omega_{\rm de0}$.
This specifies the constant $\cM^4$ from Eq.(\ref{Omega-phi0-M4}),
noticing that $\bar{\tilde\chi}_0\bar{K}'_0=\sqrt{\bar{\tilde\chi}_0} U_0$ from the
definition (\ref{U-A-def}) and using Eq.(\ref{sqrt-chi-a}). Choosing the normalization
$\bar{K}_0=-1$ [as seen from the scalar-field Lagrangian in the action
(\ref{S-def}), we must choose a normalization of the kinetic function $K$ to
specify the scale $\cM^4$], this yields
\be
\frac{\cM^4}{\bar\rho_0} = \frac{\Omega_{\varphi 0}}{\Omega_{\rm m0}}
+ \frac{\epsilon_{2,0}}{-3\epsilon_{2,0} + \left. \frac{d\ln U}{d\ln a} \right|_{a=1} }
\;\;\; \mbox{with} \;\;\; \bar{K}_0 = - 1 ,
\label{M4-Omega0-K0-def}
\ee
and we recover the fact that $\cM^4 \simeq \bar\rho_{\rm de 0}$.

\subsubsection{The reconstruction mapping}

Once the energy scale $\cM^4$ has been obtained from the cosmological parameters,
using Eqs.(\ref{Omega-phi-0-def}) and (\ref{M4-Omega0-K0-def}),
and the values of the given functions $\epsilon_2$ and $d\ln U/d\ln a$ today,
we directly obtain $\bar{\tilde\chi}(a)$ from Eq.(\ref{sqrt-chi-a}).
Next, from $\bar{\tilde\chi}(a)$ and the definition of $U(a)$ in Eq.(\ref{U-A-def}) we
obtain $\bar{K}'(a)$,
\be
\bar {K}' (a)= \frac{U(a)}{a^3 \sqrt{\bar{\tilde\chi}(a)}}.
\ee
This also gives $K(a)$ from the integration of
\be
\bar K(a)= -1 + \int_0^{\ln a} \frac{U(a)}{a^3 \sqrt{\bar{\tilde\chi}(a)}} \,
\frac{d\bar{\tilde\chi}}{d\ln a}(a) \ d\ln a ,
\ee
with the initial condition $\bar{K}(a=1)=\bar{K}_0=-1$.
Inverting the mapping $a\to \bar{\tilde\chi}(a)$, we can reconstruct $K(\chi)$ from $U(a)$
and $\bar A(a)$.

Finally, the Friedmann equation (\ref{Friedmann-H-1}) provides $H(a)$,
as a function of the Hubble rate today $H_0$.
The value of the scalar field $\bar\varphi$ is obtained from the integration of
the second expression in (\ref{dphi-dt-dphi-dlna}), with the initial condition
$\bar\varphi_0=0$ (which has no physical impact)
\be
\varphi(a)=  - \int_0^{\ln a} \frac{\sqrt{2\bar{\tilde\chi}\cM^4}}{\bar{A}H} \ d\ln a.
\ee
The curves $\{\bar{\tilde\chi}(a),\bar{K}(a)\}$ and $\{\bar\varphi(a),\bar{A}(a)\}$
provide a reconstruction of the kinetic and coupling functions
$K(\tilde\chi)$ and $A(\varphi)$, over the range probed by the cosmological
background. This is a tomographic reconstruction mapping as the whole theory can be reconstructed from the knowledge of $U(a)$ and $\bar A (a)$ as functions of the scale factor.

\subsection{Properties of the characteristic functions $U(a)$ and $\bar{A}(a)$}
\label{sec:Properties-U-A}

\subsubsection{Cosmological evolution}

The functions $U(a)$ and $\bar{A}(a)$ that define the K-mouflage model
through the parameterization described in the previous section cannot be
fully arbitrary. They must obey several constraints that are associated with
the properties (\ref{signs-convention})-(\ref{W+-cond}).

First, the Klein-Gordon equation (\ref{KG-1}) can actually be integrated as
\cite{Brax:2014wla,Brax:2015lra}
\be
\bar{A}^{-3} U(a) = \bar{A}^{-3} a^3 \sqrt{\bar{\tilde\chi}} \bar{K'}
= \int_0^t dt \, \frac{\beta\bar\rho_0}{\tilde{M}_{\rm Pl}\sqrt{2\cM^4}} ,
\label{KG-int1}
\ee
that is, the integration constant must vanish in order to obtain a realistic early-time
cosmology.
Indeed, if the integration constant $C$ of the Klein-Gordon equation does not vanish
we obtain $\bar{\tilde\chi} \sim t^{-4/(2m-1)}$, where $K(\tilde\chi) \sim \tilde\chi^m$ with
$m \geq 1$ in the highly non-linear regime, and $\bar\rho_{\varphi} \sim t^{-4m/(2m-1)}$.
This leads to a scalar energy density that grows faster in the past than the matter density.
Thus, recovering the $\Lambda$-CDM model at early times implies that $C=0$
\cite{Brax:2014wla}.
Then, at early time we have $d\bar\varphi/dt\rightarrow -\infty$ but $\bar\varphi$ goes
to a finite limit, which implies that $\bar{A}$ and $\beta$ also go to
constants (this also means that the Einstein and Jordan frames become equivalent
at the background level at high redshifts), and we must have $U(a) \propto t$.
This gives
\be
a \ll 1: \;\;\;  U(a) \propto a^2 \;\; \mbox{in the radiation era and} \;\;
U(a) \propto a^{3/2} \;\; \mbox{in the early matter era.}
\label{U-a-0}
\ee
Next, since we focus on simple models where the various fields are monotonic
functions of time, hence $\bar{\tilde\chi}$ is a monotonic decreasing function of
$a$, the condition (\ref{W+-cond}) implies
\be
U(a)/a^3 \;\; \mbox{is a monotonic decreasing function of} \;\; a .
\label{U-a-Wp}
\ee
The condition $U(a)/a^3\rightarrow+\infty$ for $a\rightarrow 0$ is already
implied by the constraint (\ref{U-a-0}).
From the definition (\ref{U-A-def}) and the relation (\ref{KG-2}), using $\epsilon_2<0$,
we obtain
\be
U > 0 , \;\;\; U/\bar{A}^3 \;\; \mbox{is a strictly increasing function of} \;\; a .
\label{U-A3-a}
\ee

The function $\bar{A}(a)$, and its logarithmic derivative $\epsilon_2(a)$, must obey
the constraints (\ref{signs-convention}) and (\ref{a-0-A-Kp}).
At early times we obtain from Eq.(\ref{sqrt-chi-a})
$\sqrt{\bar{\tilde\chi}} \sim -\epsilon_2/U$.
Since we look for a monotonic decreasing function $\bar{\tilde\chi}(a)$, with
$\bar{\tilde\chi}\rightarrow+\infty$ for $a\rightarrow 0$, we obtain
\be
a \ll 1 : \;\;\;  -\epsilon_2/U \;\; \mbox{is a decreasing function of} \;\; a , \;\;\;
-\epsilon_2/U \rightarrow +\infty \;\; \mbox{for} \;\; a \rightarrow 0 .
\label{epsilon2-U-a}
\ee
This means that $\epsilon_2$ cannot go to zero too fast at high redshift.
From (\ref{U-a-0}) we can see that $\epsilon_2$ converges to zero more slowly
than $a^{3/2}$ in the early matter era and than $a^2$ in the radiation era.

At late times, the K-mouflage scenario leads to a $\Lambda$-CDM like accelerated
expansion, driven by a constant dark energy density. The scalar field $\bar\varphi$ and
the conformal factor $\bar{A}$ converge to constant values, $t \sim \ln(a)$
and $U(a) \sim \ln(a)$. Therefore, far in the dark energy era, we have
\be
a \gg 1 : \;\; U(a) \propto \ln(a) .
\label{U-lna}
\ee
This asymptotic behaviour applies to the future; the current epoch, where dark
energy is only starting to dominate the Universe, only sees the beginning  of this
logarithmic dependence.
However, this means that over the range $0<a<1$ the function $U(a)$ should first
grow as $a^2$, and then as $a^{3/2}$, from (\ref{U-a-0}), and further slow down
at $a \lesssim 1$ to eventually reach the logarithmic growth of (\ref{U-lna}).
In particular, choosing functions $U(a)$ that are strictly increasing functions of the
scale factor automatically ensures that the condition (\ref{U-A3-a}) is satisfied,
because $\bar{A}(a)$ is monotonically decreasing ($\epsilon_2<0$).
For conformal factors $\bar{A}$ and strength $\beta(a)$ that do not vary much
over the cosmological evolution, Eq.(\ref{KG-int1}) actually shows that $U(a)$ grows
as $U(a) \sim t(a)$ where $t(a)$ is the cosmic time, and this should be close to the
$\Lambda$-CDM expansion.

\subsubsection{Normalising U(a)}

As shown in section~\ref{sec:Building-parameterization}, the parameterization of the
K-mouflage model in terms of $U(a)$ and $\bar{A}(a)$ allows us to normalize the
cosmology to the present cosmological parameters,
$\{\Omega_{\rm m0},\Omega_{\rm rad0},\Omega_{\rm de0},H_0\}$.
In addition, the function $\bar{A}$ is normalized to unity today and goes to a finite
value $A_*>1$ at high redshift, with $|\epsilon_2| \lesssim 0.01$ to satisfy Solar
System constraints,
\be
\bar{A}(0)=A_* > 1 , \;\;\; \bar{A}(1)=1 , \;\;\; \epsilon_2<0, \;\;\; | \epsilon_2 |
\lesssim 0.01 .
\label{A-epsilon2}
\ee
This also means that $\bar{A}(a)$ is always close to unity, with typically
$(A_*-A) \lesssim 0.1$.

When we define the K-mouflage model from its kinetic function $K(\tilde\chi)$, as in the
action (\ref{S-def}), we usually impose the low-$\chi$ expansion
$K(\tilde\chi) = -1 + \tilde\chi + ...$,
where the dots stand for higher-order terms, which sets the normalization of the model
without any loss of generality.
The factor $-1$ defines the normalization of the energy scale parameter $\cM^4$.
The normalization to unity of the linear term defines the normalization of the
scalar field $\varphi$ and of the kinetic factor $\tilde{\chi}$.
We have already performed the first normalization in Eq.(\ref{M4-Omega0-K0-def})
in the case of the $\{U(a),\bar{A}(a)\}$ parameterization,
except that we choose $\bar{K}(a=1)=-1$ instead of $\bar{K}(a\rightarrow+\infty)=-1$.
Thus, we still have one degeneracy associated with the second normalization, which we
have not performed yet.
Indeed, going through the reconstruction method described in
section~\ref{sec:Building-parameterization}, we can check that the cosmology
and our relations are invariant through the transformation
$U \rightarrow \lambda U$, $\tilde\chi \rightarrow \tilde\chi/\lambda^2$,
$K' \rightarrow \lambda^2 K'$, $K \rightarrow K$, $\varphi \rightarrow \varphi/\lambda$.
Therefore, we need to choose the normalization of $U(a)$ today, to fully characterize
the model.
Using this symmetry we normalize $U(a)$ by
\be
\bar{K}'_0 = 1 , \;\;\; \mbox{whence} \;\;\;
U_0 = \sqrt{\frac{- \bar\rho_0 \epsilon_{2,0}}{{\cal M}^4 2 (-3 \epsilon_{2,0}
+ \left. \frac{d\ln U}{d\ln a}\right|_{a=1} )} } ,
\label{norm-U-Kp}
\ee
where the second equation comes from the combination of Eqs.(\ref{U-A-def}) and
(\ref{sqrt-chi-a}).
Thus, the normalization of the kinetic function $K(\tilde\chi)$ is not exactly identical to
the one used in previous works
\cite{Brax:2014wla,Brax:2014yla,Barreira:2014gwa,Brax:2015lra}.
Instead of $K(\tilde\chi=0)=-1$ and
$K'(\tilde\chi=0)=1$ we now have $K(\tilde\chi=\bar{\tilde\chi}_0)=-1$ and
$K'(\tilde\chi=\bar{\tilde\chi}_0)=1$. However, in practice we
have $\bar{\tilde\chi}_0 \ll 1$, as can be seen from Eq.(\ref{KG-int1}) which yields
$\bar{\tilde\chi}_0 \sim \beta^2$ and we must have $\beta \lesssim 0.1$ to satisfy
Solar System and cosmological constraints \cite{Barreira:2015aea}.
Therefore, these two different normalizations are usually rather close.

Finally, the standard $\Lambda$-CDM cosmology is recovered in the limit
$\epsilon_2\rightarrow 0$, as $U \sim \sqrt{|\epsilon_2|}$ with the normalization
choice (\ref{norm-U-Kp}), $\bar{\tilde\chi} \sim |\epsilon_2|$, $\bar{K}' \sim 1$,
$|\bar{K}-\bar{K}_0| \sim |\epsilon_2|$, $\bar\varphi \sim \sqrt{|\epsilon_2|}$,
$\beta \sim \sqrt{ |\epsilon_2| }$ and $|\bar{A}-1| \sim |\epsilon_2|$.

\subsection{What does $U(a)$ represent?}
\label{sec:power-law-kinetic}

It is interesting to see more precisely how the shape of the functions $U(a)$ and
$\bar{A}(a)$ are related to the shape of the kinetic function $K(\tilde\chi)$.
For a power-law large-$\tilde\chi$ behaviour,
\be
\tilde\chi \gg 1: \;\; K(\tilde\chi) \sim \tilde\chi^m \;\;\; \mbox{with} \;\;\; m > 1 ,
\label{K-m-def}
\ee
we have in the early matter era
\be
\mbox{early matter era:} \;\;\; \bar{\tilde\chi} \sim a^{-3/(2m-1)} , \;\;\;
U \sim a^{3/2} \sim t , \;\;\; \epsilon_2 \sim - a^{3(m-1)/(2m-1)} ,
\label{m-early-matter}
\ee
and in the radiation era
\be
\mbox{radiation era:} \;\;\; \bar{\tilde\chi} \sim a^{-2/(2m-1)} , \;\;\;
U \sim a^{2} \sim t , \;\;\; \epsilon_2 \sim - a^{(4m-3)/(2m-1)} .
\label{m-radiation}
\ee
We can check that these behaviours satisfy the constraint (\ref{epsilon2-U-a}) over
the full range $1 \leq m<+\infty$.
We recover the behaviours (\ref{U-a-0}) and we find that the function $U(a)$ is largely
independent of the kinetic function $K(\tilde\chi)$. In agreement with Eq.(\ref{KG-int1}),
it is mostly related to the expansion history of the universe as
\be
U(a) \sim t(a)
\ee
and should not deviate much from the ``$\Lambda$-CDM shape'' $U_{\Lambda\rm-CDM}(a)$
that is obtained by setting $\bar{A}$ and $\beta$ to constants in Eq.(\ref{KG-int1}),
unless one considers coupling functions $A(\varphi)$ with steep slopes.
The main freedom, associated with the shape of $K(\tilde\chi)$, is contained in the
function $\epsilon_2(a)$.
In particular, its power-law exponent in  the scale factor,
$\epsilon_2 \sim - a^{\nu_A}$, is directly related to the exponent of the kinetic function,
with $\nu_A=3(m-1)/(2m-1)$ in the early matter era.

\subsection{Factors $\epsilon_1$ and $\epsilon_2$}
\label{sec:Factors-epsilon}

In addition to the factor $\epsilon_2$ defined in (\ref{Friedmann-2}), a key
factor that enters the growth of large-scale structures is the function
$\epsilon_1$ defined in (\ref{Poisson-K}), which enters the evolution equation
(\ref{D-linear-Jordan}) of the linear growing modes of the density contrast
and describes the amplification of gravity in the unscreened regime.

The function $\epsilon_2(a)$ is directly obtained from the function $\bar{A}(a)$
within the parameterization $\{U(a),\bar{A}(a)\}$, as
$\epsilon_2=d\ln\bar{A}/d\ln a$.
It is interesting to see how the characteristic function $\epsilon_1(a)$ is related
to this parameterization.
From the last Eq.(\ref{dphi-dt-dphi-dlna}) and the definition
(\ref{Poisson-K}) of $\epsilon_1$,
we have $\epsilon_1=\epsilon_2^2\bar{A}^2H^2\tilde{M}_{\rm Pl}^2
/\bar{\tilde\chi}\bar{K}'\cM^4$.
On the other hand, from the definition (\ref{U-A-def}) we have
$\bar{\tilde\chi}\bar{K}'=\sqrt{\bar{\tilde\chi}}U/a^3$ and from Eq.(\ref{sqrt-chi-a})
we obtain $\bar{\tilde\chi}\bar{K}'=-\bar{\rho}_0\epsilon_2\bar{A}^4
/2\cM^4a^3(-3\epsilon_2+d\ln U/d\ln a)$.
Combining these two results gives
\be
\epsilon_1(a) = - \, \epsilon_2 \, \frac{2 \left( -3\epsilon_2+\frac{d\ln U}{d\ln a} \right)}
{3\Omega_{\rm m}} .
\label{eps1-eps2}
\ee
In the matter dominated era we have $\Omega_{\rm m}\simeq 1$ and
$d\ln U/d\ln a \simeq 3/2$, see (\ref{U-a-0}). This yields
\be
\mbox{matter era:} \;\;\; \epsilon_1 \simeq -\epsilon_2 .
\label{eps1-eps2-matter-era}
\ee
At late time in the dark energy era the ratio $\epsilon_1/|\epsilon_2|$ grows
because $\Omega_{\rm m}$ decreases in the denominator of (\ref{eps1-eps2}).
Therefore, in the K-mouflage models the functions $\epsilon_1$ and $\epsilon_2$
are equal for most of the cosmological evolution and measure both the drift
of Newton's constant (or the Jordan-frame Planck mass) and the amplification
of the gravitational force in the growth of large-scale structures.

\section{Numerical results}
\label{sec:Numerical}

We now present our numerical results for several K-mouflage models defined by the
parameterization $\{U(a),\bar{A}(a)\}$ described in
section~\ref{sec:Parameterizing-K-mouflage}.

\subsection{Explicit models}
\label{sec:explicit}

In the following sections, we describe our reconstruction procedure and our analysis
with a simple choice for the functions $U(a)$ and $\bar{A}(a)$, with a few parameters.
We shall consider three sets of parameters that cover different behaviours for the
underlying K-mouflage models, within the constraints that must satisfied.
The precise forms of the functions (\ref{A-param-def}) and (\ref{U-param-def}) have
no particular theoretical meaning. They are only the simplest choices that satisfy
the previous requirements and are sufficient for our illustration purposes.

For the function $\bar{A}(a)$ we take
\be
\gamma_A  > 0 : \;\;\;  \bar{A}(a) = 1 + \alpha_A - \alpha_A
\left[ \frac{(\gamma_A+1) a}{\gamma_A+a} \right]^{\nu_A} .
\label{A-param-def}
\ee
This is the simplest function that goes to constants at both early and late
times, $a\rightarrow 0$ and $a\rightarrow \infty$, with a power-law behaviour $a^{\nu_A}$
at early times. The parameters $\alpha_A$, $\gamma_A$ and $\nu_A$ are related to the
more physical quantities $\epsilon_{2,0}$ and $m$ by
\be
\epsilon_{2,0} = - \frac{\alpha_A \gamma_A \nu_A}{\gamma_A+1} , \;\;\;
\nu_A= \frac{3(m-1)}{2m-1} \;\; \mbox{with} \;\; m > 1 ,
\label{nuA-def}
\ee
where $m$ is the exponent of the kinetic function in the matter era regime,
defined by $K \sim \tilde\chi^m$, see (\ref{m-early-matter}).
If we require the kinetic function to keep the same exponent in $\tilde\chi$
in the radiation era we should change $\nu_A$ for $a<a_{\rm eq}$, following
(\ref{m-radiation}). However, because the dark energy component is negligible
in the radiation era and cannot be measured at these high redshifts (for the models
that we consider here) we keep the same form (\ref{A-param-def}) with a single
parameter $\nu_A$ given by Eq.(\ref{nuA-def}) at all redshifts.
Thus, the parameterization (\ref{A-param-def}) is determined by the three parameters
$\{m,\epsilon_{2,0},\gamma_A\}$, $\alpha_A$ being obtained from
Eq.(\ref{nuA-def}).
As explained above, $m$ is the logarithmic slope of the kinetic function
at large $\tilde\chi$, $\epsilon_{2,0}=d\ln \bar{A}/d\ln a$ at $a=1$ today,
and $\gamma_A \sim 1$ describes the transition to the dark energy epoch, where
$\bar{A}$ converges to a finite value as in (\ref{chi-phi-A-infty}).
In particular, we have
 \be
\bar{A}(a=0)=1+\alpha_A,\ \bar{A}(a=1)=1, \
\bar{A}(a=\infty)=1+\alpha_A-\alpha_A (\gamma_A+1)^{\nu_A}.
\ee
The parameter $\epsilon_{2,0}$ measures the deviation of $\bar{A}$ from unity and
the departure of the K-mouflage model from the $\Lambda$-CDM cosmology.

For the function $U(a)$ we consider the shape
\be
\alpha_U > 0 , \;\;\; \gamma_U \geq 1 : \;\;\; U(a) \propto \frac{a^2 \ln(\gamma_U+a)}
{(\sqrt{a_{\rm eq}}+\sqrt{a}) \ln(\gamma_U+a) + \alpha_U a^2} ,
\label{U-param-def}
\ee
which comes from the simple combination
\be
\frac{1}{U} \propto \frac{\sqrt{a_{\rm eq}}}{a^2} + \frac{1}{a^{3/2}}
+ \frac{\alpha_U}{\ln(\gamma_U+a)} ,
\label{U-inv-param-def}
\ee
where $a_{\rm eq}$ is the scale factor at equality between the matter and radiation
components.
This is the simplest function that follows the evolution of $t(a)$ through
the radiation, matter and dark-energy eras, each term successively dominating in
Eq.(\ref{U-inv-param-def}). In agreement with the requirements (\ref{U-a-0}) and
(\ref{U-lna}), it gives $U(a) \sim a^2$ first, followed by
$U(a)\sim a^{3/2}$ and next $U(a)\sim \ln a$.
The transition between the radiation and matter eras is set by $a_{\rm eq}$ whereas
the two parameters $\alpha_U$ and $\gamma_U$ describe the transition to the dark
energy era and are of order unity.
Finally, the normalization of $U(a)$ is obtained from Eq.(\ref{norm-U-Kp}).

\begin{table}[tbp]
\centering
\begin{tabular}{|c|cc|ccc|}
\hline
model &  $\alpha_U$ & $\gamma_U$ & $m$ & $\epsilon_{2,0}$ & $\gamma_A$ \\
\hline
1 & 0.2 & 1 & 3 & -0.01 & 0.2 \\
2 & 0.2 & 1 & 3 & -0.01 & 0.5 \\
3 & 0.4 & 1 & 3 & -0.01 & 2 \\
\hline
\end{tabular}
\caption{\label{tab:models} Parameters for the three K-mouflage models
studied in section~\ref{sec:Numerical}.}
\end{table}

In the following, we present our numerical results for the three K-mouflage models
defined by the parameters given in Table~\ref{tab:models}.
They have $\epsilon_2=-0.01$, which ensures that Solar system tests and
cosmological bounds are satisfied, while other parameters are of order unity.
We keep $\gamma_U$ and $m$ unchanged, because the results are less sensitive
to these parameters, and we vary $\gamma_A$ and $\alpha_U$.
The main parameter that determines the cosmological behaviours is $\gamma_A$.
Since $m=3$, all three models satisfy a cubic form, $K \sim \tilde\chi^3$, in the highly
non-linear regime.
However, as we shall find below, these three sets of parameters correspond
to significantly different kinetic functions $K(\tilde\chi)$ and allow us to explore a wide
range of K-mouflage behaviours.

For the cosmological parameters we choose the Planck mission results,
$\Omega_{\rm m0}=0.3175$, $\Omega_{\rm de0}=0.6825$ and $H_0=67.11$ km/s/Mpc.

\subsection{Reconstruction of the kinetic and coupling functions}
\label{sec:Reconstruction}

\subsubsection{Evolution with time}
\label{sec:Evolution}

\begin{figure}
\begin{center}
\epsfxsize=7.5 cm \epsfysize=5.8 cm {\epsfbox{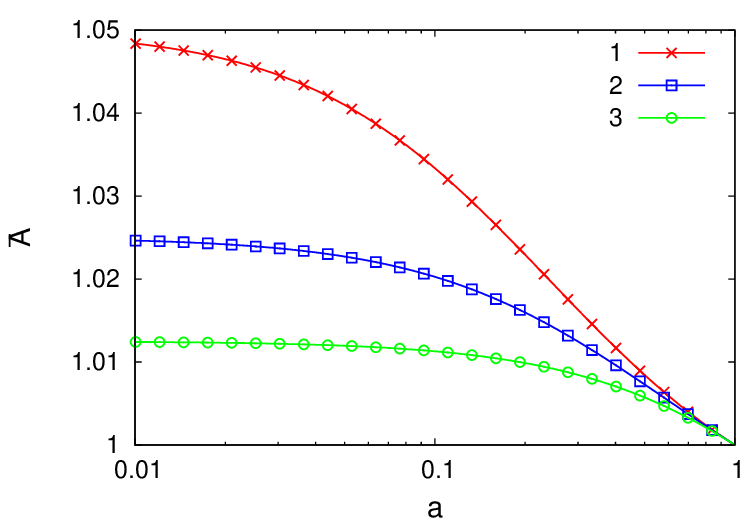}}
\epsfxsize=7.5 cm \epsfysize=5.8 cm {\epsfbox{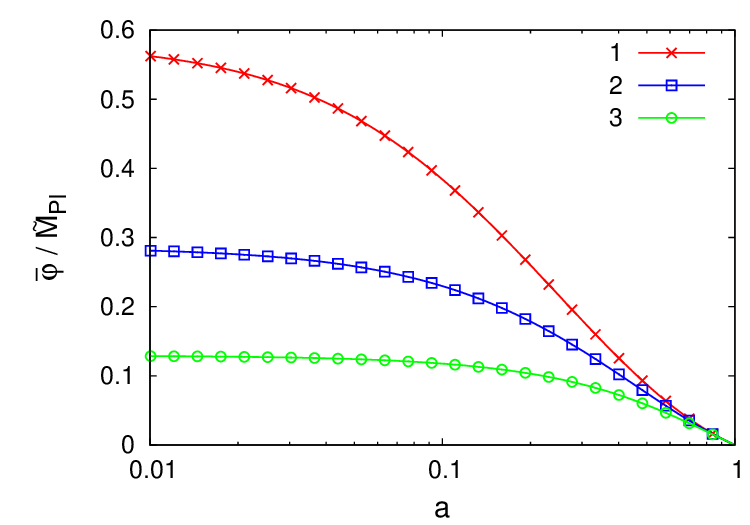}}
\end{center}
\caption{
{\it Left panel:} background conformal factor $\bar{A}(a)$ as a function of the scale
factor, for the three K-mouflage models defined in Table~\ref{tab:models} where
$\gamma_A$ decreases from bottom to top.
{\it Right panel:} background scalar field in units of the Einstein-frame Planck mass,
$\bar\varphi/\tM_{\rm Pl}$, as a function of the scale factor.
}
\label{fig_A_phi_a}
\end{figure}

\begin{figure}
\begin{center}
\epsfxsize=7.5 cm \epsfysize=5.8 cm {\epsfbox{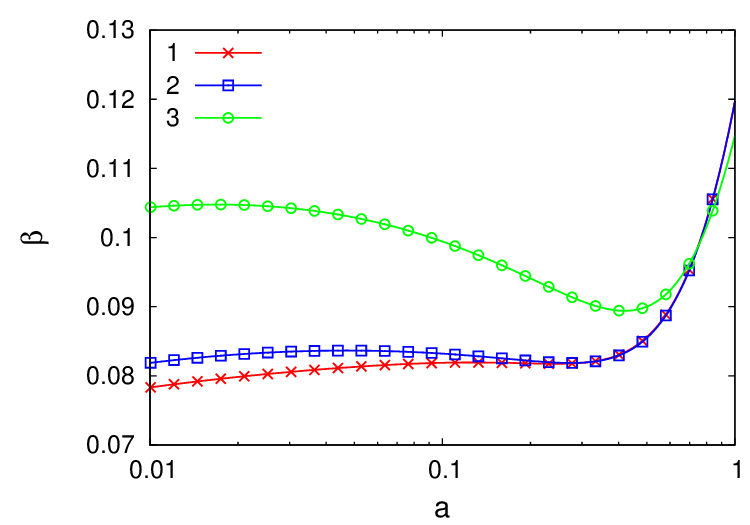}}
\epsfxsize=7.5 cm \epsfysize=5.8 cm {\epsfbox{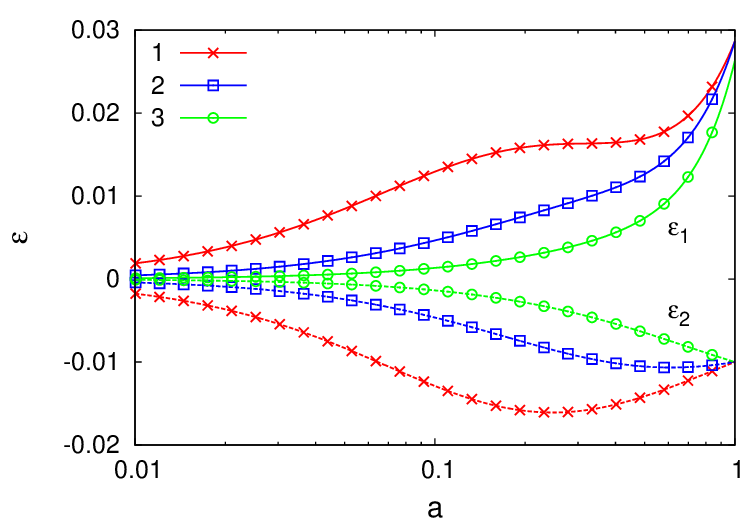}}
\end{center}
\caption{
{\it Left panel:} coupling strength
$\beta(a)=\tilde{M}_{\rm Pl}d\ln\bar{A}/d\bar\varphi$.
{\it Right panel:} functions $\epsilon_1(a)=2\beta^2/\bar{K}'$ (upper solid curves)
and $\epsilon_2(a)=d\ln\bar{A}/d\ln a$ (lower dashed curves).
}
\label{fig_beta_eps_a}
\end{figure}

\begin{figure}
\begin{center}
\epsfxsize=7.5 cm \epsfysize=5.8 cm {\epsfbox{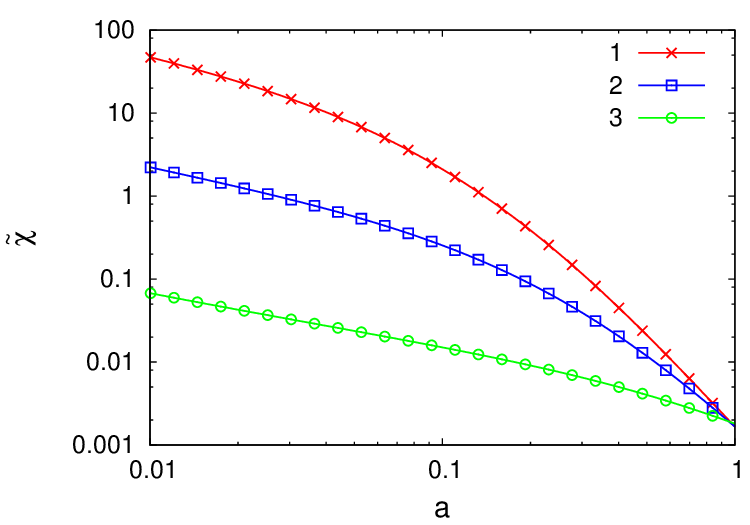}}
\epsfxsize=7.5 cm \epsfysize=5.8 cm {\epsfbox{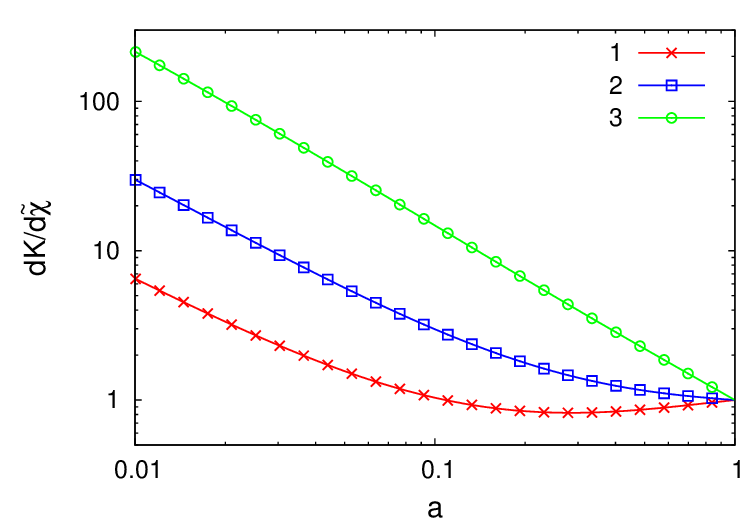}}
\end{center}
\caption{
{\it Left panel:} background kinetic factor $\bar{\tilde\chi}$.
{\it Right panel:} first derivative of the kinetic function,
$\bar{K}'=d\bar{K}/d\bar{\tilde\chi}$.
}
\label{fig_chi_Kp_a}
\end{figure}

From the functions $U(a)$ and $\bar{A}(a)$ and the cosmological parameters today,
we can compute the expansion history of the Universe, as described in
section~\ref{sec:Building-parameterization}.
We show our results for the three K-mouflage models defined in Table~\ref{tab:models}
in Figs.~\ref{fig_A_phi_a} - \ref{fig_chi_Kp_a}.

We can see in the left panel of Fig.~\ref{fig_A_phi_a} that as $\gamma_A$ decreases,
from models 3 to 1, the deviation of the coupling function $\bar{A}$ from unity
grows, in agreement with Eq.(\ref{A-param-def}) and $\bar{A}(a=0)=1+\alpha_A$.
This will lead to a greater deviation from the $\Lambda$-CDM cosmology,
as the fifth force arises from the variations of the conformal coupling factor
$A$.

As seen in the right panel of Fig.~\ref{fig_A_phi_a}, a greater cosmological range for
$\bar{A}(a)$ also leads to a greater excursion for the field $\bar\varphi$.
Indeed, from the definition of $\beta$ and $\epsilon_2$ we have
$d\bar\varphi/d\ln a = \tilde{M}_{\rm Pl} \epsilon_2/\beta$.
Therefore, if $\beta$ does not vary too much, a greater range of $\bar{A}$,
whence a greater $|\epsilon_2|$ at intermediate redshifts, leads to a larger time
derivative of the scalar field.
Thus, as could be expected, a greater range for the coupling function $A(\bar\varphi)$
corresponds to a greater range for the scalar field $\bar\varphi$ itself.

At low $z$, both $\bar{A}(a)$ and $\bar\varphi(a)$ converge for all three models.
This is because we normalize all models to the same cosmological parameters
and to the same slope $\epsilon_{2,0}=-0.01$ of $\bar{A}(a)$ today.

As seen in the left panel of Fig.~\ref{fig_beta_eps_a}, $\beta$ remains
about $0.1$ for all three models and at all redshifts.
Indeed, we typically have $|\epsilon_2| \sim \epsilon_1 = 2\beta^2/\bar{K}'$,
see Eqs.(\ref{eps1-eps2}) and (\ref{eps1-eps2-matter-era}). At late times we also
have $\bar{\tilde\chi} \simeq 0$ and $\bar{K}' \simeq 1$, which gives
$|\epsilon_2| \sim \epsilon_1 \simeq 2\beta^2$. Therefore, $|\epsilon_{2,0}| = 0.01$
also implies $\beta_0 \sim 0.1$.
Then, $\beta$ remains almost constant with redshift because we chose a
function $U(a)$ that behaves like $t(a)$, see Eqs.(\ref{KG-int1})
and (\ref{U-a-0}).
This ensures a coupling function $A(\varphi)$ that is smooth and ``natural'',
without sharp knees or oscillations that would introduce new scales.

We show the two factors $\epsilon_1$ and $\epsilon_2$ in the right panel of
Fig.~\ref{fig_beta_eps_a}. In agreement with Eqs.(\ref{eps1-eps2}) and
(\ref{eps1-eps2-matter-era}), $\epsilon_1 \simeq -\epsilon_2$ at high $z$
and $\epsilon_1/|\epsilon_2|$ grows at low $z$ as $1/\Omega_{\rm m}$.
The factors $\beta$, $\epsilon_1$ and $\epsilon_2$ converge at low $z$
for all three models, because we normalize the models at $z=0$.
For the model 1, which shows the greater deviation of $\bar{A}$ from unity
at high $z$, the amplitude of $\epsilon_2$ slightly grows from $z=0$ to $z\simeq 4$.
At early time, $a\rightarrow 0$, $\epsilon_1$ and $\epsilon_2$ converge to zero
for all models.
For $\epsilon_2$ this is ensured by our choice of the coupling function
$\bar{A}(a)$, with $\epsilon_2 \sim - a^{\nu_A}$ from Eq.(\ref{A-param-def}).
Since at early times we have $\epsilon_1 \simeq -\epsilon_2$ this also implies
$\epsilon_1 \sim a^{\nu_A}$. From the definition of $\epsilon_1$
in (\ref{Poisson-K}) this also means that $\bar{K}' \rightarrow +\infty$ at high $z$.
The fact that $\epsilon_1$ is of order $1\%$ implies that deviations of the linear
power spectrum from the $\Lambda$-CDM prediction will be of the order of a few
percents, see Eq.(\ref{D-linear-Jordan}).

Finally, we show the kinetic terms $\bar{\tilde\chi}$ and $\bar{K}'$
in Fig.~\ref{fig_chi_Kp_a}.
Again, both $\bar{\tilde\chi}$ and $\bar{K}'$ converge at low $z$
for all three models.
We can check that $\bar{\tilde\chi}$ and $\bar{K}'$ increase at high $z$.
When $\bar{K}' \gg 1$ the K-mouflage mechanism comes into play with the dominance
of non-linear effects in the K-mouflage Lagrangian. This cosmological
K-mouflage screening ensures that the dark energy component becomes subdominant
as compared with the matter density. However, in contrast with the constant
dark energy density shown by the $\Lambda$-CDM cosmology, $\bar\rho_{\rm de}$
grows with $z$.
We find that the smaller values of $\gamma_A$, associated with larger
$\bar{A}$ and $\bar\varphi$, correspond to greater $\bar{\tilde\chi}$ and
smaller $\bar{K}'$.
Indeed, at fixed $U(a)$ we can see from Eq.(\ref{U-A-def}) that
$\bar{K}' \sim 1/\sqrt{\bar{\tilde\chi}}$, so that $\bar{K}'$ must decrease if
$\bar{\tilde\chi}$ increases.
Next, Eq.(\ref{sqrt-chi-a}) gives $\sqrt{\bar{\tilde\chi}} \sim -\epsilon_2$
so that $\sqrt{\bar{\tilde\chi}}$ increases with $|\epsilon_2|$.

For the model 1 we find that $\bar{K}'$ slightly decreases from $z=0$ to $z\simeq 4$.
This will lead to a large deviation from the $\Lambda$-CDM matter clustering,
as the factor $\epsilon_1$ defined in Eq.(\ref{Poisson-K}) remains
unsuppressed by the denominator $\bar{K}'$ until $z\simeq 4$.
This can also be seen in the upper curve of the right panel in
Fig.~\ref{fig_beta_eps_a}.
However, one cannot take $\gamma_A$ too small to achieve low values of
$\bar{K}'$ at $z \sim 2$. Indeed, to have a well behaved K-mouflage scenario,
without ghosts and small scale instabilities, we must have $K' >0$ and
$K'+2\tilde\chi K''>0$.
Moreover, the reconstruction developed in section~\ref{sec:Building-parameterization}
breaks down if $\bar{K}'$ becomes negative, since a positive function $U(a)$ cannot
be consistent with a negative $\bar{K}'$ in Eq.(\ref{U-A-def}).
These constraints define a lower bound on the parameter $\gamma_A$, for given
values of the other parameters.

\subsubsection{Reconstructed coupling and kinetic functions}
\label{sec:Reconstructed}

\begin{figure}
\begin{center}
\epsfxsize=7.5 cm \epsfysize=5.8 cm {\epsfbox{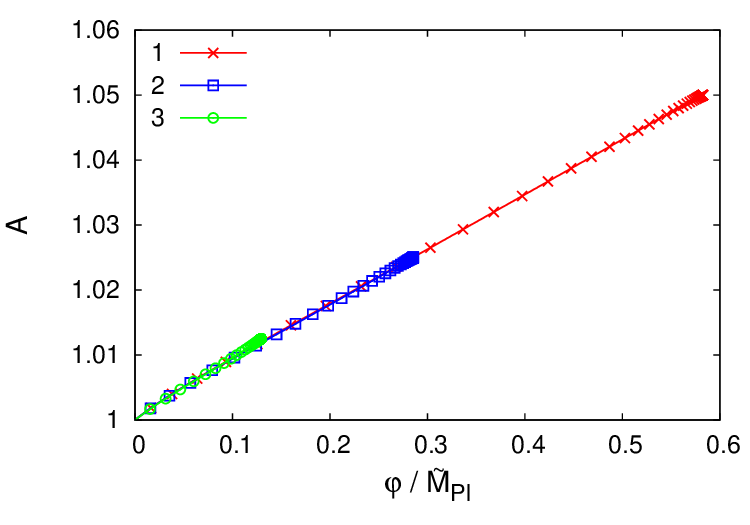}}
\epsfxsize=7.5 cm \epsfysize=5.8 cm {\epsfbox{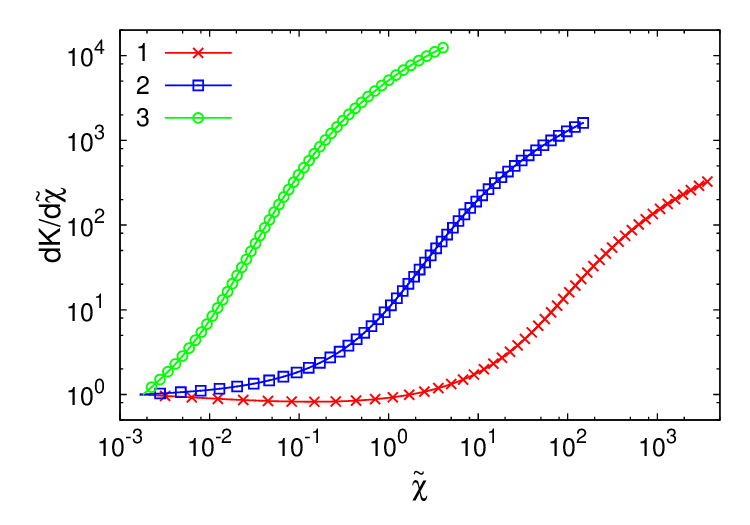}}
\end{center}
\caption{
{\it Left panel:} reconstructed coupling function $A(\varphi)$.
{\it Right panel:} reconstructed first derivative of the kinetic function,
$K'(\tilde{\chi})=dK/d\tilde{\chi}$.
}
\label{fig_Aphi_Kpchi}
\end{figure}

The previous reconstruction of the time evolution of cubic models from the
$\{U(a),\bar{A}(a)\}$ parameterisation allows us to rebuild the field dependence
of both $A(\varphi)$ and $K'(\tilde\chi)$.
This is shown in Fig.~\ref{fig_Aphi_Kpchi}.
The coupling function $A(\varphi)$ is essentially linear over the range of background
field values $\bar\varphi$ probed by the models in the past of the Universe.
This is consistent with the fact that its first derivative $\beta$ is almost constant
and close to $0.1$, as seen in the left panel of Fig~\ref{fig_beta_eps_a}.
As explained in section~\ref{sec:Evolution}, this is due to the facts that we considered
smooth functions $U(a)$ and $\bar{A}(a)$ and that the scalar field excursions are
not large, $\bar\varphi/\tilde{M}_{\rm Pl} \lesssim 0.6$, as seen in the right panel
in Fig.~\ref{fig_A_phi_a}.

On the other hand, the reconstruction of $K'(\tilde\chi)$ clearly shows that the kinetic
function is very sensitive to the value of the parameter $\gamma_A$.
Because of our common normalization of the models at $z=0$,
$\bar{\tilde\chi}(a)$ and $\bar{K}'(a)$ converge at low $z$ to common values for
all three models, as seen in Fig.~\ref{fig_chi_Kp_a}.
This implies that $K'(\tilde\chi) = 1$ for the common value
$\bar{\tilde\chi}_0 \simeq 2\times 10^{-3}$ today.
At high values of $\tilde{\chi}$ in the early matter era, the kinetic function reaches
a cubic form, whence $K' \sim \tilde{\chi}^2$. This transition happens at much lower
redshifts for the model 3, with a high $\gamma_A$, than for the model 1, with
a low $\gamma_A$.
This is consistent with Fig.~\ref{fig_chi_Kp_a}, where we found that a smaller
$\gamma_A$, whence a greater $|\epsilon_2|$ at intermediate redshifts,
leads to a greater $\bar{\tilde\chi} \sim |\epsilon_2|^2$ and a lower
$\bar{K}' \sim 1/|\epsilon_2|$. Therefore, as we decrease $\gamma_A$ the
curve $K'(\tilde\chi)$ shifts to the lower right corner in Fig.~\ref{fig_Aphi_Kpchi}.
In particular, the almost constant value $\bar{K}'(a) \simeq 1$ until $z\simeq 10$
found in the right panel in Fig.~\ref{fig_chi_Kp_a} for model 1 translates into
a plateau with $K' \simeq 1$ until $\tilde{\chi} \simeq 10$ in
Fig.~\ref{fig_Aphi_Kpchi}.
This gives a model where there is no K-mouflage screening of the cosmological
background until $z \simeq 10$.
At very high values of $\tilde\chi$ we can see a change of slope of the kinetic function.
This is because we enter the radiation era, where the relation between the
exponent $\nu_A$ of $\epsilon_2(a)$ and the exponent $m$ of $K(\tilde\chi)$
changes from (\ref{m-early-matter}) to (\ref{m-radiation}).
With our choice of a fixed $\nu_A$ set by the matter era relation,
$m=3$ and $\nu_A=3(m-1)/(2m-1)=6/5$, we find in the radiation era
$m_{\rm rad}=9/8$ so that $K' \sim \tilde\chi^{1/8}$ for
$\tilde\chi > \bar{\tilde\chi}_{\rm eq}$, where $\bar{\tilde\chi}_{\rm eq}$ is the
value reached at equality between the matter and radiation components.
Since the K-mouflage model is very close to a $\Lambda$-CDM cosmology at these
high redshifts we only reconstructed the kinetic function $K'(\tilde\chi)$
up to $z = 10^4$ in Fig.~\ref{fig_Aphi_Kpchi}.

\subsection{Comparison with $\Lambda$-CDM}
\label{sec:comparison}

\begin{figure}
\begin{center}
\epsfxsize=5.0 cm \epsfysize=5.6 cm {\epsfbox{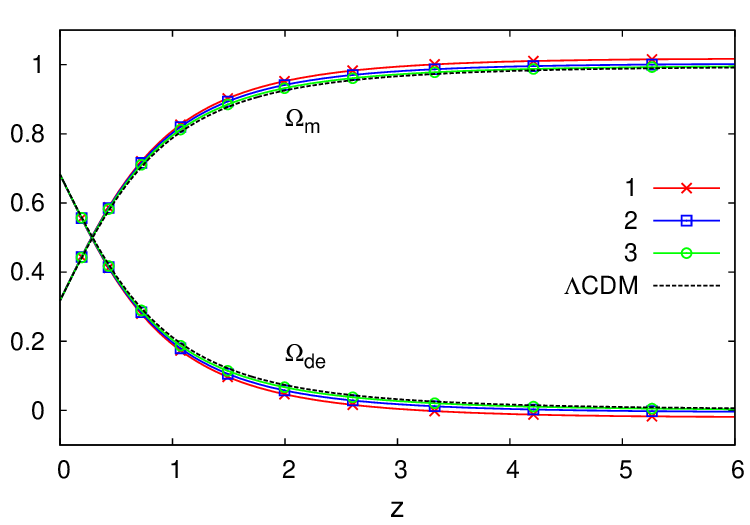}}
\epsfxsize=5.0 cm \epsfysize=5.6 cm {\epsfbox{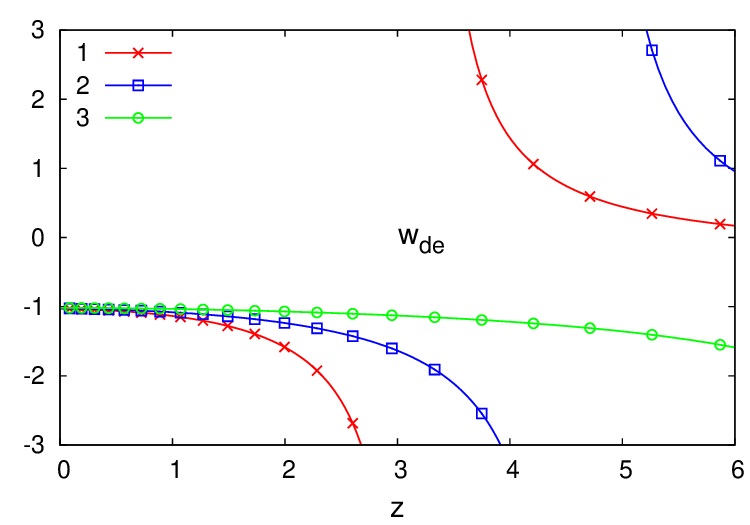}}
\epsfxsize=5.0 cm \epsfysize=5.6 cm {\epsfbox{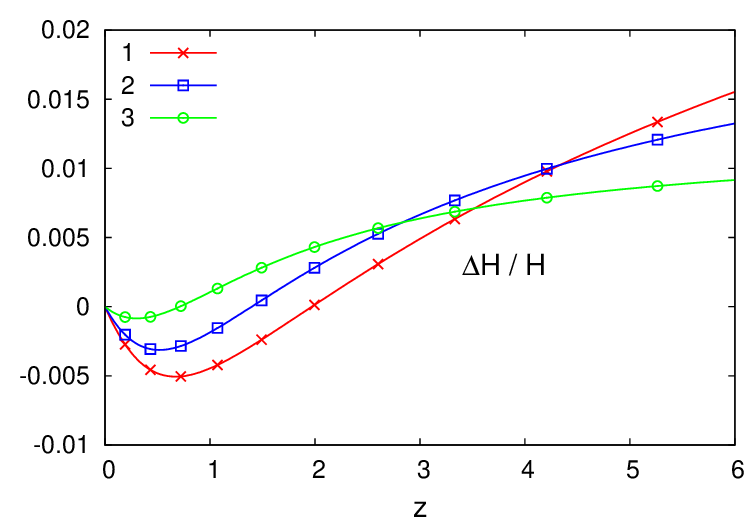}}
\end{center}
\caption{
{\it Left panel:} matter and dark energy cosmological density parameters,
as a function of the redshift. We show the three K-mouflage models of
Table~\ref{tab:models} and the $\Lambda$-CDM reference cosmology.
{\it Middle panel:} effective dark energy equation of state parameter
$w_{\rm de}=\bar{p}_{\rm de}/\bar{\rho}_{\rm de}$ for the K-mouflage models.
{\it Right panel:} relative deviation of the Hubble rate from the $\Lambda$-CDM reference,
for the three K-mouflage models.
}
\label{fig_Om_H_z}
\end{figure}

\begin{figure}
\begin{center}
\epsfxsize=5 cm \epsfysize=5.6 cm {\epsfbox{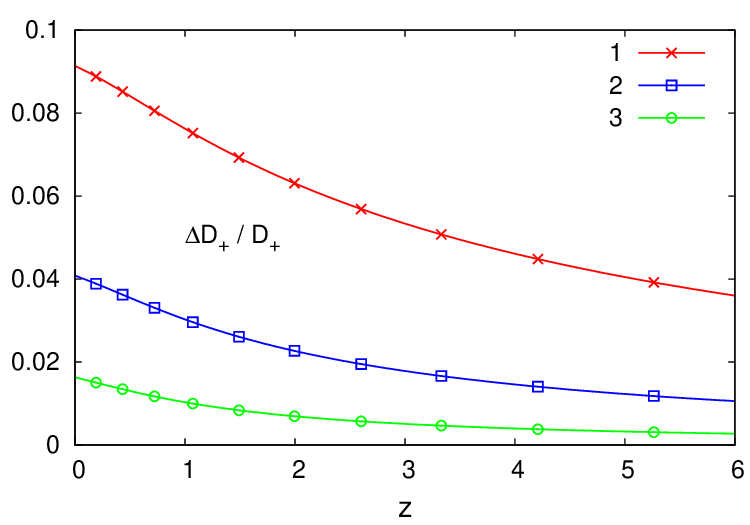}}
\epsfxsize=5 cm \epsfysize=5.6 cm {\epsfbox{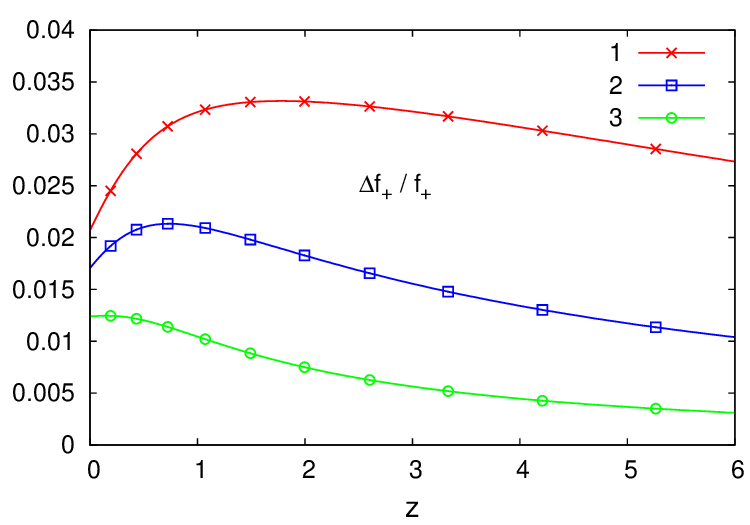}}
\epsfxsize=5 cm \epsfysize=5.6 cm {\epsfbox{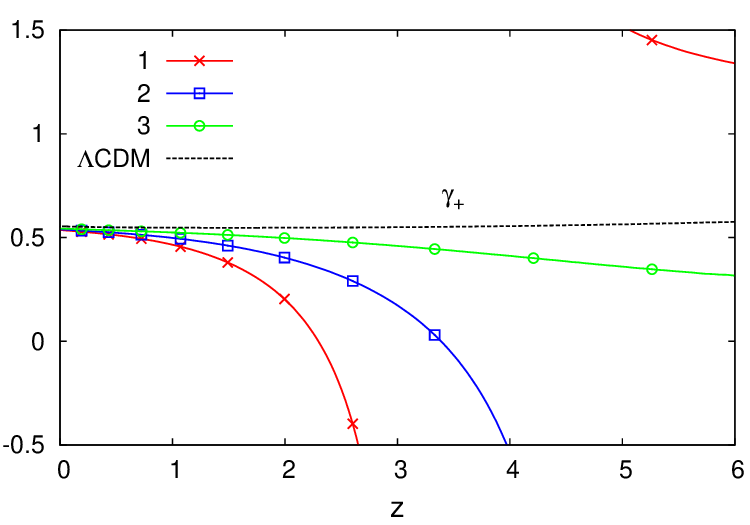}}
\end{center}
\caption{
{\it Left panel:} relative deviation of the linear growing mode $D_+(z)$ from the
$\Lambda$-CDM reference.
{\it Middle panel:} relative deviation from the $\Lambda$-CDM reference of the
linear growth rate $f_+=d\ln D_+/d\ln a$.
{\it Right panel:} growth rate parameter $\gamma_+(z)$ for the K-mouflage models and
the $\Lambda$-CDM reference.
}
\label{fig_Dlin_f_z}
\end{figure}

The K-mouflage models that we have reconstructed in the previous section show deviations
from $\Lambda$-CDM at  both the background and perturbative levels.
We compare the K-mouflage results to the $\Lambda$-CDM predictions in
Figs.~\ref{fig_Om_H_z} - \ref{fig_Dlin_f_z}, where all scenarios obey the same
normalization for the cosmological parameters today.

We can check in the left panel in Fig.~\ref{fig_Om_H_z} that the cosmological
density parameters $\Omega_{\rm m}(z)$ and $\Omega_{\rm de}(z)$ remain close
to the $\Lambda$-CDM predictions, with percent deviations as for other background
quantities. This order of magnitude is set by the value of $\epsilon_{2,0}$.
In agreement with previous works, the dark energy density becomes negative at high
redshift in the K-mouflage scenarios, which yields $\Omega_{\rm m}>1$ and
$\Omega_{\rm de}<0$. However, this only occurs at high $z$ when the dark energy
density is subdominant. Therefore this feature is not a problem as $H^2>0$ always.
This change of sign of $\bar{\rho}_{\rm de}$ corresponds to the divergence and
change of sign of the effective dark energy equation of state parameter
$w_{\rm de}=\bar{p}_{\rm de}/\bar{\rho}_{\rm de}$ shown in the middle panel
in Fig.~\ref{fig_Om_H_z}. At low $z$, when the dark energy density becomes important,
all models converge to $w_{\rm de} \simeq -1$ as we remain close to the
$\Lambda$-CDM cosmology. Notice that the dark energy equation of state can cross the phantom divide $w_{\rm de}<-1$ even though the propagation speed
$c_s^2>0$, as shown in the right panel in Fig.~\ref{fig_M24_cs2_a} below.
This is a drastic difference with $K$-essence models \cite{Vikman:2004dc} due to the coupling to matter of K-mouflage.
We can check in the right panel in Fig.~\ref{fig_Om_H_z} that the Hubble rates converge
to the same normalization $H_0$ today. As we move to higher redshift,
the K-mouflage Hubble rates first grow more slowly than in the $\Lambda$-CDM
scenario, which yields a negative deviation $\Delta H$ at $z\sim 1$. This is due to the
decrease of the K-mouflage dark energy density with redshift until it vanishes and next
becomes negative, as explained above, whereas the $\Lambda$-CDM dark energy
density remains constant (the matter densities remain identical, $\bar{\rho}_0/a^3$,
in all scenarios).
The K-mouflage Hubble rates become greater than the $\Lambda$-CDM expansion rate
at high $z$, when the dark energy density is negligible.
This is due to the time dependence of the K-mouflage Planck mass (in the Jordan frame).
Indeed, all Planck masses are normalized to the same value $\tilde{M}_{\rm Pl}$ today,
and we have seen in Eq.(\ref{Lambda-c-J-def}) that the K-mouflage Planck mass
is given by $M_{\rm Pl}^2=f \tilde{M}_{\rm Pl}^2 = \bar{A}^{-2} \tilde{M}_{\rm Pl}^2$.
This gives a Newton's constant ${\cal G}_{\rm N} = \tilde{\cal G}_{\rm N} \bar{A}^2$
that grows with redshift and leads to a greater Hubble expansion rate from the Friedman
equation.
The time dependence of the K-mouflage Newton constant can be read from
the time dependence of the conformal factor $\bar{A}$ shown in the left panel in
Fig.~\ref{fig_A_phi_a}. For instance, in the model 1 with the greatest deviation from
$\Lambda$-CDM the Newton constant is higher by $10\%$ at high redshift,
$z \gtrsim 100$, than its value today.

In agreement with expectations, the deviations from the $\Lambda$-CDM background
quantities decrease from model 1 to 2 and from 2 to 3.
This corresponds to smaller values of the factors $\epsilon_1$ and $\epsilon_2$
and to a higher kinetic function derivative $\bar{K}'$, that is, to a stronger
K-mouflage non-linear screening of the cosmological background.

From the reconstruction of the K-mouflage model, and more precisely from its
Hubble expansion rate and the factor $\epsilon_1$ defined in Eq.(\ref{Poisson-K}),
we can obtain the linear growing and decaying modes $D_{\pm}(a)$ of the matter
density field from Eq.(\ref{D-linear-Jordan}).
This also provides the linear growth rate $f_+(a)$ and the growth rate parameter
$\gamma_+(a)$ defined by
\be
f_+(a) \equiv \frac{d\ln D_+}{d\ln a} , \;\;\;
\gamma_+(a) = \frac{\ln f_+}{\ln\Omega_{\rm m}} ,
\label{linear-rate-def}
\ee
with $f_+ = \Omega_{\rm m}^{\gamma_+}$.

We show the relative deviations of $D_+(z)$ and $f_+(z)$ from the
$\Lambda$-CDM predictions in Fig.~\ref{fig_Dlin_f_z}.
The deviation of the linear growing mode $D_+$ can reach up to $9\%$ at $z=0$,
even though $\epsilon_2=-0.01$ today. This is because the impact of the modified
gravity is cumulative with time on the linear growing mode. This can yield a significant
amplification factor, as compared with the magnitude of $\epsilon_2$,
if K-mouflage screening is delayed to high redshift, as for model 1.
The deviations of the growth rate $f_+$ are of the order of a few percents.
Again, as for the background quantities, the deviations of the perturbed quantities
from the $\Lambda$-CDM predictions increase from model 3 to 2 and from 2 to 1.
This corresponds to higher $\epsilon_1$ and $\epsilon_2$ and lower $\bar{K}'$.
In terms of the parameters of Table~\ref{tab:models} this corresponds to
a lower $\gamma_A$.

The right panel in Fig.~\ref{fig_Dlin_f_z} shows that the behaviour of the growth rate
parameter $\gamma_+(z)$ obtained in the K-mouflage models is very different from
the one found in the $\Lambda$-CDM cosmology.
This is because of two effects: a) the fifth force amplifies the growth of structures and
b) the dark energy density actually becomes negative at high redshift
in the K-mouflage scenarios, which corresponds to $\Omega_{\rm m}>1$ and to
a change of sign of $\ln\Omega_{\rm m}$.
At low $z$, as in $\Lambda$-CDM scenario, we have $f_+ <1$ and
$\Omega_{\rm m} < 1$, and because we always remain close to the $\Lambda$-CDM
cosmology we have $\gamma_+ \simeq 0.55$. As we go to higher redshift,
we converge to the Einstein-de Sitter cosmology where both $f_+$
and $\Omega_{\rm m}$ go to unity. Therefore, both $\ln f_+$ and $\ln\Omega_{\rm m}$
become very sensitive to the details of the cosmological model and their sign
can change with the models. In the $\Lambda$-CDM case we always have
$f_+<1$ and $\Omega_{\rm m}<1$, whence $\gamma_+>0$, and the parameter
$\gamma_+$ does not evolve much.
In contrast, in the K-mouflage scenarios we have $f_+>1$ at high redshift because
of the acceleration of structure formation by the fifth force. This leads to a change
of sign of $\gamma_+$, which becomes negative at $z \simeq 2.3$ for model 1,
and at higher redshifts for models 2 and 3.
Next, at still higher redshift ($z \gtrsim 3$ for model 1) the dark energy density
becomes negative while $\Omega_{\rm m}$ becomes greater than unity.
This gives rise to a second change of sign of $\gamma_+$, which turns positive
again, after it diverges by going from $-\infty$ to $+\infty$.
The behaviour of $\gamma_+(z)$ is thus a distinct signature of these K-mouflage
scenarios.
However, there is no dramatic change to the linear growing mode $D_+$ nor to the
linear growth rate $f_+$ and this only relies on the change of sign of $\ln f_+$
and $\ln\Omega_{\rm m}$ while $f_+$ and $\Omega_{\rm m}$ are close to unity
in the matter dominated era.
In practice, it is probably more robust to measure and use independently
$\Omega_{\rm m}(z)$ and $f_+(z)$ in the data analysis, as the exponent $\gamma_+$
is likely to be rather unstable because of this high sensitivity to small deviations in the
matter era.
We leave to future works a detailed analysis of these points.

\subsection{Effective time dependent functions}
\label{sec:effective-functions}

\begin{figure}
\begin{center}
\epsfxsize=7.5 cm \epsfysize=5.8 cm {\epsfbox{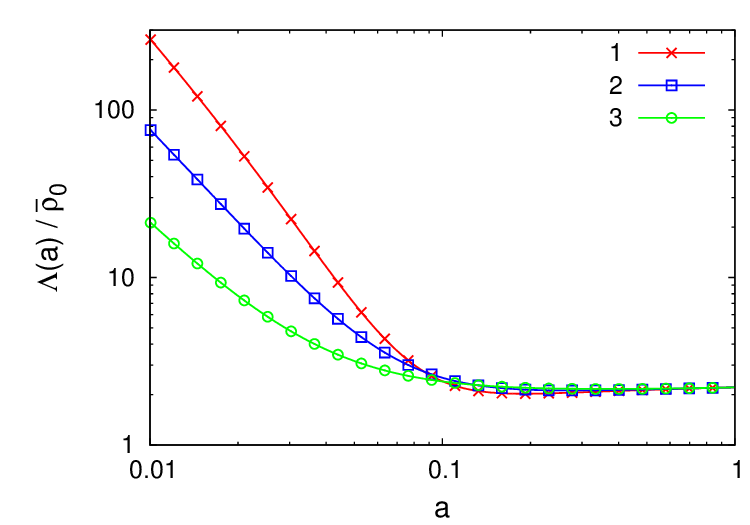}}
\epsfxsize=7.5 cm \epsfysize=5.8 cm {\epsfbox{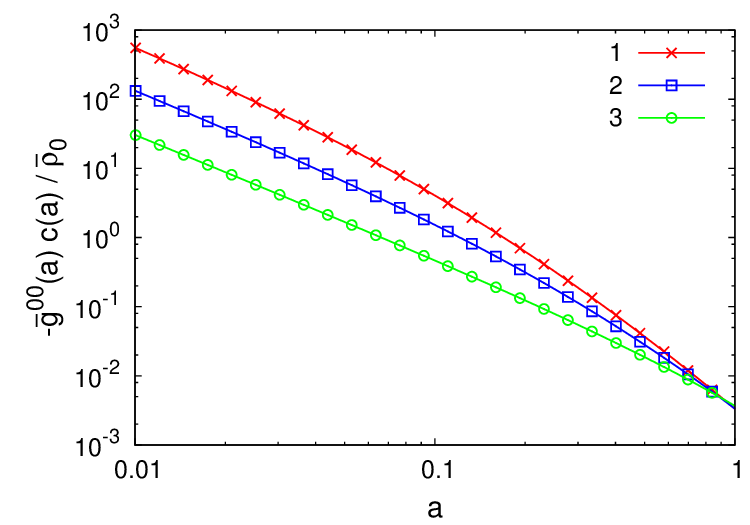}}
\end{center}
\caption{
{\it Left panel:} effective function $\Lambda(a)$ in units of $\bar\rho_0$.
{\it Right panel:} effective function $c(a)$, multiplied by the factor
$-\bar{g}^{00}/\bar\rho_0$.
}
\label{fig_Lambda_c_a}
\end{figure}

\begin{figure}
\begin{center}
\epsfxsize=7.5 cm \epsfysize=5.8 cm {\epsfbox{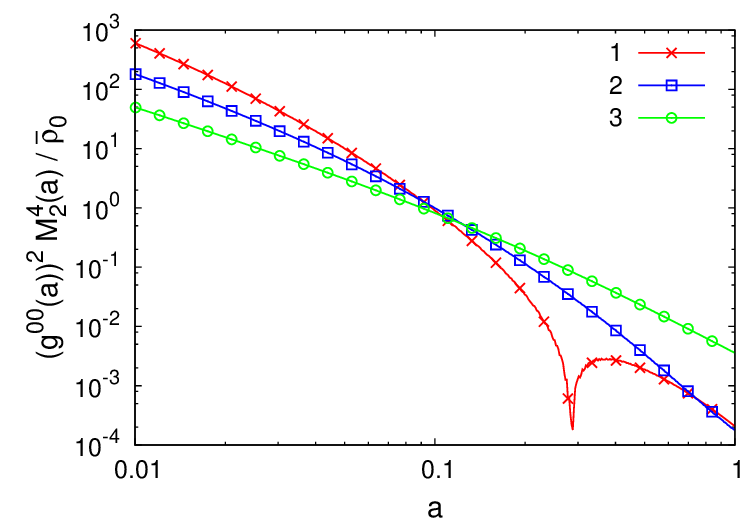}}
\epsfxsize=7.5 cm \epsfysize=5.8 cm {\epsfbox{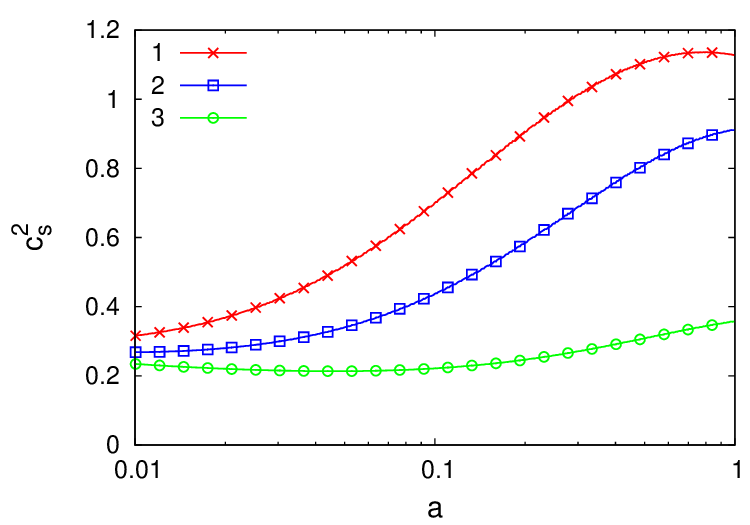}}
\end{center}
\caption{
{\it Left panel:} effective function $M_2^4(a)$ multiplied by the factor
$(\bar{g}^{00})^2/\bar\rho_0$.
{\it Right panel:} scalar propagation speed $c_s^2$ (in units of the speed of light).
Notice that it is always positive.
}
\label{fig_M24_cs2_a}
\end{figure}

The effective action depends on several operators with coupling functions which are time
dependent. For the case of K-mouflage scenarios they can be expressed in terms
of the K-mouflage kinetic and coupling functions $K(\tilde\chi)$ and $A(\varphi)$
as in Eqs.(\ref{f-Lambda-c-Mn-Einstein}) and (\ref{Lambda-c-J-def}), which give
\be
\begin{split}
f(a) = \bar{A}^{-2} , \;\;\; M_{\rm Pl}^2(a) = \bar{A}^{-2} \tilde{M}_{\rm Pl} , \;\;\;
\Lambda(a) = \bar{A}^{-4} \cM^4 ( \tilde{\chi} \bar{K}' - \bar{K} ) , \\
- \bar{g}^{00}(a) c(a) = \bar{A}^{-4} \cM^4 \tilde{\chi} \bar{K}'
- \frac{\bar\rho}{\Omega_{\rm m}} \epsilon_2^2 , \;\;\;
(\bar{g}^{00}(a))^2 M_2^4(a) = \bar{A}^{-4} \cM^4 \tilde{\chi}^2 \bar{K}'' .
\end{split}
\label{EFT-functions}
\ee
We show these effective functions in Figs.~\ref{fig_Lambda_c_a} and
\ref{fig_M24_cs2_a}, as well as the propagation speed $c_s^2$, given by
\cite{Garriga:1999vw}
\be
c_s^2 = \frac{\bar K'}{\bar K' +2\bar{\tilde\chi} \bar K''} .
\label{cs2-def}
\ee

The function $\Lambda (a)$ generalises the cosmological constant to this
dynamical setting while $c(a)$ represents the effects of the kinetic energy of
the K-mouflage field.
In Fig.~\ref{fig_Lambda_c_a}, we find that the cosmological evolution of
$\Lambda (a)$ only
starts at $z \sim 10$ when dark energy becomes dynamical. This corresponds to the
time when $- \bar{g}^{00}(a) c(a)$ becomes of order unity and $K(\tilde\chi)$ starts
to deviate from the low-$\tilde\chi$ constant regime $\bar{K} \simeq -1$.
In the early matter era $\Lambda(a)$ and $- \bar{g}^{00}(a) c(a)$ grow with redshift
as $\bar{K} \sim \bar{\tilde\chi} \bar{K}' \sim \bar{\tilde\chi}^m$.
The difference between the behaviours of $\Lambda(a)$ and $- \bar{g}^{00}(a) c(a)$
at low redshifts comes from the fact that $\Lambda(a)$ is sensitive to the
non-zero value of the kinetic function at $\tilde\chi=0$, which plays the role of the
cosmological constant, whereas $c(a)$ is only sensitive to the kinetic term
$\bar{K}' \bar{\tilde\chi}$ and decays as $\bar{\tilde\chi}$ at late times.
Again, the deviations from the $\Lambda$-CDM behaviour, a constant $\Lambda(a)$,
are greater for the model 1, which corresponds to smaller $\gamma_A$.

At high redshift the quadratic mass term $(\bar{g}^{00})^2 M_2^4$ shows the same
growth with $z$ as $\Lambda(a)$ and $- \bar{g}^{00}(a) c(a)$, as
$\bar{\tilde\chi}^m$, for a kinetic function that has a power law asymptote at high
$\tilde\chi$.
For the model 1 this term actually becomes negative at low redshift, with the change of
sign of $\bar{K}''$, in agreement with the reconstructed kinetic function displayed in
Fig.~\ref{fig_Aphi_Kpchi} [we can see in that figure that $K'(\tilde\chi)$ has a negative
slope over $10^{-3} \lesssim \tilde\chi \lesssim 0.5$].
This implies that at low redshift the scalar propagation speed $c_s$ is greater than the
speed of light for this model, as we can check in the right panel in Fig.~\ref{fig_M24_cs2_a}.
At high redshift, and at all times for the other models where $K'(\tilde\chi)$ is a monotonic
increasing function, $c_s$ is smaller than the speed of light.

This again illustrates the different behaviours obtained for K-essence and K-mouflage
models. Whereas in K-essence scenarios $c_s^2$ has to be greater than unity in some
redshift range in order to reproduce a $\Lambda$-CDM-like expansion history
\cite{Bonvin:2006}, in K-mouflage scenarios we usually have $c_s^2<1$ around the
cosmological background at all redshifts, except for a subclass of models
such that $K''<0$ at low $\chi$ which have $c_s^2>1$ at low redshifts.
These different behaviours come from the fact that the evolution equations of the scalar field
are different. In the K-essence scenario the Klein-Gordon equation has no external source
term and shows several fixed points, which also depend on the current expansion era.
Then, the evolution of the Universe (as it goes from radiation, matter and dark-energy eras)
drives the scalar field from one fixed point to another \cite{ArmendarizPicon:2000dh}.
In the K-mouflage scenario, the scalar field is driven by the coupling to matter, which gives
rise to an external source term in the Klein-Gordon equation (\ref{KG-1}).
Another important difference is that the dark-energy equation-of-state parameter
$w_{\rm de}$ of Eq.(\ref{w-Jordan-def}) is different from the scalar-field equation-of-state
parameter $w_{\varphi}$ of Eq.(\ref{wphi-def}).
This leads to different behaviours for $w_{\rm de} (z)$ and $c_s^2(z)$.

The superluminality encountered in the case of model 1 in the late-time Universe
does not entail any causality issue in the present Universe in astrophysical situations
where Newton's potential is small enough, see the appendix \ref{sec:causality}
for more details.

\section{Conclusion}
\label{sec:Conclusion}

We have described how K-mouflage models can be expressed within the
framework of the effective field theory of dark energy. Because these K-mouflage scenarios
are fully defined by their nonlinear Lagrangian, we obtain a fully nonlinear
effective-field-theory action that can be expanded up to all orders.
In the unitary gauge,
we find that the K-mouflage scenarios only generate the operator
$(\delta g^{00}_{(u)})^n$ at each order $n$, whereas more general scenarios can
also generate additional operators involving the extrinsic curvature of the constant time slicing surface.
This is a distinct signature of K-mouflage models within the class of models coupled to matter \footnote{$K$-essence would also have the same type of expansion but its features
are drastically different, for instance the slip parameter $\gamma$ (see appendix A.2)) would only differ from one only for K-mouflage.}. If observations analysed
within the effective field theory framework show that such additional operators
are required to match data, this would at once rule out all these K-mouflage
models.

At the quadratic order, which fully determines the cosmological background and
the linear regime of the growth of large-scale structures,
the effective action is defined in terms of time dependent functions,
$f(\tau)$, $\Lambda(\tau)$, $c(\tau)$ and $M_2^4(\tau)$, and we have shown
how they can be related to the K-mouflage kinetic and coupling functions
$K(\tilde\chi)$ and $A(\varphi)$.
We have also described how the usual equations of motion associated with
K-mouflage models are recovered within the effective field theory approach.
In particular, at the linear level the scalar field $\pi$ that appears within this
formalism is related to the original K-mouflage scalar field $\varphi$
by a relation of the form $\pi \propto \delta\varphi$.
However, the two fields are not identical.
Although we obtain a closed form for the exact nonlinear equation of motion of the
field $\pi$, it has a significantly more complex form than the one of the original
field $\varphi$.

Next, we have presented a simple parameterisation, in terms of two
time dependent functions $U(a)$ and $\bar{A}(a)$, that fully determines the
cosmological K-mouflage behaviour.
One advantage of this parameterisation, as compared with the definition in terms
of $K(\tilde\chi)$ and $A(\varphi)$, is that it provides a simple reconstruction
of the past cosmological behaviour from the normalization of the cosmological
parameters today.
In contrast, given generic functions $K(\tilde\chi)$ and $A(\varphi)$, because the
equations are non-linear one must apply some iterative search algorithm
to set the mass scale $\cM^4$ in order to reach a given dark energy cosmological
parameter today (which means computing the full background evolution from
$a\ll 1$ until today for each trial).

By explicitly choosing the function $\bar{A}(a)$, and therefore $\epsilon_2(a)$,
we also ensure that some cosmological and Solar System constraints are satisfied.
More precisely, $|\epsilon_{2,0}| \leq 0.01$ ensures that the current drift of
Newton's constant remains below the upper bound obtained from the
Lunar Ranging experiment.
Moreover, we typically have $|\epsilon_2| \sim \epsilon_1 = 2\beta^2/\bar{K}'$,
see Eqs.(\ref{eps1-eps2}) and (\ref{Poisson-K}). At late times we also
have $\bar{\tilde\chi} \simeq 0$ and $\bar{K}' \simeq 1$, which gives
$|\epsilon_2| \sim \epsilon_1 \simeq 2\beta^2$. Therefore, $|\epsilon_{2,0}| \leq 0.01$
also implies $\beta \lesssim 0.1$. This ensures that the upper bound
on the deviation from the Hubble expansion rate predicted by the $\Lambda$-CDM
cosmology at Big Bang Nucleosynthesis is satisfied (this deviation is not due to the dark energy component,
which is subdominant at high redshift, but to the time dependence of the Planck
mass). It also means that deviations from the $\Lambda$-CDM matter power spectrum
are of the order of a few percents on large linear scales, which remains consistent
with observations.

By choosing an exponent $m>1$ for the highly non-linear regime of the kinetic
function, $K(\tilde\chi) \sim \tilde\chi^m$, which corresponds to
$\epsilon_2(a) \sim - a^{(4m-3)/(2m-1)}$ in the matter era within the
parameterization $\{U,\bar{A}\}$, we also ensure that the dark energy component
is subdominant at high $z$ and we recover the Einstein-de Sitter expansion in the
matter era. Then, all cosmological constraints are satisfied at these qualitative
and semi-quantitative orders. A more precise study, through a combined MCMC
analysis using various cosmological probes, would provide the exact parameter
space of K-mouflage models that is allowed in terms of the parameters of
Table~\ref{tab:models}.

However, we must point out that the analysis presented in this paper does not
tackle all Solar System constraints. In particular, the shape of the kinetic
function $K(\tilde\chi)$ at large negative $\tilde\chi$ is strongly constrained
and must satisfy $K' \gg 1$ and $|\tilde\chi K''| \ll K'$.
This is beyond the reach of the effective field theory approach, which can only
probe the kinetic function $K(\tilde\chi)$ in the cosmological regime
$\tilde\chi>0$.
Thus, the K-mouflage scenarios provide a simple example where the
cosmological and Solar System regimes are mostly decoupled, although
there are some common constraints as explained above, such as the value of the
coupling strengh $\beta$.

The results presented in this paper can be useful for numerical studies of
cosmological probes at the background and linear levels.
In particular, numerical codes that have been designed for the effective field theory
framework can use the functions
$f(\tau)$, $\Lambda(\tau)$, $c(\tau)$ and $M_2^4(\tau)$ obtained in this paper
to compute the observational signatures of K-mouflage models. This is left for future work.

\acknowledgments
We would like to thank N. Bartolo and M. Liguori for discussions and suggestions.
This work is supported in part by the French Agence Nationale de la Recherche under Grant ANR-12-BS05-0002. P.B.
acknowledges partial support from the European Union FP7 ITN
INVISIBLES (Marie Curie Actions, PITN- GA-2011- 289442) and from the Agence Nationale de la Recherche under contract ANR 2010
BLANC 0413 01.

\appendix

\section{Effective-field-theory action at quadratic order}
\label{app:S2-def}

\subsection{Scalar dynamics}
\label{app:Scalar-dynamics}

The unitary gauge is useful to determine the operators which define the effective action of
K-mouflage. In particular, the action (\ref{S-unit2}) shows that only the operator
$\delta g^{00}_{(u)}$ is generated by the K-mouflage models.
On the other hand, it hides the presence of the scalar degree of freedom which modifies
gravity and affects the growth of structure.
As recalled in section~\ref{sec:scalar-field-pi}, the scalar field can be unravelled
by the transformation (\ref{Stueckelberg}), which restores the Lorentz invariance
through the explicit introduction of the scalar field $\pi$.

In section~\ref{sec:scalar-field-pi} we considered the fully nonlinear longitudinal
gauge action (\ref{S-J-unit1}), which led to the exact action (\ref{S-pi-J}).
However, in practice one uses effective field theories to describe theories defined
up to some finite perturbative orders, including all operators that are allowed by
symmetries up to that order. This has the benefit of providing a very general framework,
which can encompass many different theories, at the cost of leaving the nonlinear
closure unspecified.
In this spirit, if we only consider the equations of motion at linear order over the
field fluctuations, we only need the effective action up to quadratic order
over the metric potentials $\Phi$ and $\Psi$ defined in
Eq.(\ref{ds2-Newtonian-gauge}) and over the scalar field $\pi$. Notice that
$\tau$ coincides with the time-slicing of the background cosmology.
In the small-scale and non-relativistic limits considered in this paper, this gives
\be
\begin{split}
S_2 = & \int d^4x \; a^2 \biggl \lbrace  f \tM_{\rm Pl}^2 \biggl [ - 3 \dot{\Psi}^2 + (\nabla\Psi)^2
- 2 \nabla\Phi\cdot\nabla\Psi + \frac{\dot{f}}{f} \left( 3 \dot{\pi} \dot{\Psi}
+ \nabla \pi \cdot \nabla (\Phi - 2 \Psi) \right)  \\
&  + \frac{3}{4} \left( \frac{\dot{f}}{f} \right)^2
\left( - \dot{\pi}^2 + (\nabla\pi)^2 \right) \biggl ] + a^2 f^2 \cM^4 \bar{\tilde\chi}
\biggl [ (\bar{K}' + 2 \bar{\tilde\chi} \bar{K}'' ) \dot{\pi}^2 - \bar{K}' (\nabla\pi)^2 \biggl ]
- a^2 \Phi \delta\rho \biggl \rbrace ,
\end{split}
\label{S2-Jordan-1}
\ee
which also reads as
\be
\begin{split}
S_2 = & \int d^4x \; a^2 \biggl \lbrace  f \tM_{\rm Pl}^2 \biggl [ - 3 \dot{\Psi}^2 + (\nabla\Psi)^2
- 2 \nabla\Phi\cdot\nabla\Psi + \frac{\dot{f}}{f} \left( 3 \dot{\pi} \dot{\Psi}
+ \nabla \pi \cdot \nabla (\Phi - 2 \Psi) \right) \biggl ] \\
&  + c \left[ \dot{\pi}^2 - (\nabla\pi)^2 \right] + \frac{2 M_2^4}{a^2} \dot{\pi}^2
- a^2 \Phi \delta\rho \biggl \rbrace ,
\end{split}
\label{S2-Jordan-EFT-1}
\ee
where $\delta\rho$ is the pressureless matter perturbation in the Jordan frame.
In the first expression (\ref{S2-Jordan-1}) we wrote the quadratic action in terms
of the original K-mouflage kinetic function $K(\tilde\chi)$, whereas in the
second expression (\ref{S2-Jordan-EFT-1}) we used the effective-field-theory functions
$f(\tau)$, $c(\tau)$ and $M_2^4(\tau)$, which are given by Eqs.(\ref{Lambda-c-J-def})
or (\ref{EFT-functions}).
As expected, in the Jordan-frame action (\ref{S2-Jordan-1}) there is some mixing
between the scalar field $\pi$ and the gravitational potentials $\Phi$ and $\Psi$.
This corresponds to the fact that the metric potentials explicitly depend on the scalar
field $\pi$ in Eq.(\ref{Poisson-Phi-Psi-pi}).

To diagonalize the kinetic terms it is convenient to introduce the
Einstein-frame potentials,
\be
\tilde \Phi \equiv\Phi + \frac{\dot f}{2f} \pi , \;\;\;
\tilde \Psi \equiv \Psi - \frac{\dot f}{2f} \pi ,
\label{PhiE-PsiE}
\ee
and the quadratic actions (\ref{S2-Jordan-1}) and (\ref{S2-Jordan-EFT-1})
become
\be
\begin{split}
S_2 = & \int d^4x \; a^2 \biggl \lbrace f \tM_{\rm Pl}^2 \left[ - 3 \dot{\tilde\Psi}^2
+ (\nabla\tilde\Psi)^2 - 2 \nabla\tilde\Phi\cdot\nabla\tilde\Psi \right]  \\
&  + a^2 f^2 \cM^4 \bar{\tilde\chi} \left[ (\bar{K}' + 2 \bar{\tilde\chi} \bar{K}'' ) \dot{\pi}^2
- \bar{K}' (\nabla\pi)^2 \right] - a^2 \left( \tilde\Phi - \frac{\dot f}{2f} \pi \right)
\delta\rho \biggl \rbrace ,
\end{split}
\label{S2-Einstein-1}
\ee
and
\be
\begin{split}
S_2 = & \int d^4x \; a^2 \biggl \lbrace f \tM_{\rm Pl}^2 \left[ - 3 \dot{\tilde\Psi}^2
+ (\nabla\tilde\Psi)^2 - 2 \nabla\tilde\Phi\cdot\nabla\tilde\Psi \right]  \\
&  + \left( \frac{3\dot{f}^2}{4f} \tM_{\rm Pl}^2 + c \right) \left( \dot{\pi}^2 - (\nabla\pi)^2 \right)
+ \frac{2 M_2^4}{a^2} \dot{\pi}^2 - a^2 \left( \tilde\Phi - \frac{\dot f}{2f} \pi \right)
\delta\rho \biggl \rbrace ,
\end{split}
\label{S2-Einstein-EFT-1}
\ee
Using this action, we can immediately deduce that the scalar field $\pi$ couples to matter
with a strength $\dot f/2f$. Moreover the field $\pi$ propagates with a speed
$c_s$ given by Eq.(\ref{cs2-def}) in the main text,
which also reads in terms of the effective-field-theory functions as
\be
c_s^2= \frac{\frac{3 \dot{f}^2}{4 f} \tM^2_{\rm Pl} + c}
{\frac{3 \dot{f}^2}{4 f} \tM^2_{\rm Pl} + c + 2 a^{-2} M_2^4} .
\label{cs2-f-c-M2}
\ee
[Let us recall that the dot denotes a derivative with respect to conformal time
$\tau$ and we use comoving spatial coordinates.]
This expression coincides with the one obtained in \cite{Gubitosi:2012hu}
for general effective field theories, and with the one obtained in \cite{Brax:2015lra}
for K-mouflage models (in the small-scale and high-frequency limits).

\subsection{Perturbations}
\label{app:Perturbations}

We now consider the dynamics of cosmological perturbations focusing on
Cold Dark Matter. The matter energy momentum $T^{\mu\nu}$ in the Jordan frame
is conserved,
\be
\nabla_\mu T^{\mu\nu}=0 ,
\ee
implying that the continuity and the Euler equations coincide with the ones in GR.
The modifications to the dynamics of perturbations are solely induced by the change
to the Poisson equation for the Jordan-frame metric potential $\Phi$.
We focus on the sub-horizon regime of perturbations where in terms of Fourier modes
$k/a \gg H$, implying that the time variations of $\pi$ and $\tilde\Phi$ can be neglected
compared to their spatial variations.
In this quasi-static approximation, the effective action (\ref{S2-Einstein-EFT-1})
reduces to
\be
S_2 = \int d^4x \; a^2 \biggl \lbrace f \tM_{\rm Pl}^2 \left[ (\nabla \tilde\Psi)^2
- 2 \nabla\tilde\Phi \cdot \nabla\tilde\Psi \right]
- \left( \frac{3 \dot{f}^2}{4f} \tM_{\rm Pl}^2 +c \right) (\nabla\pi)^2
- a^2 \left( \tilde\Phi- \frac{\dot f}{2f} \pi \right) \delta\rho \biggl \rbrace ,
\label{S2-E-2}
\ee
where we have expressed all factors in terms of the effective-field-theory functions
$f(\tau)$ and $c(\tau)$.
The resulting Poisson equation for the Einstein-frame Newtonian potential is not modified
and we recover from the action (\ref{S2-E-2})
\be
\tilde\Phi = \tilde\Psi= \tilde\Psi_{\rm N} \;\;\; \mbox{with} \;\;\;
\frac{\nabla^2}{a^2} \tilde\Psi_{\rm N} = \frac{\delta \rho}{2M_{\rm Pl}^2} ,
\ee
where $M_{\rm Pl}^2=f \tM_{\rm Pl}^2$ is the Jordan-frame Planck mass and
$\delta\rho$ the Jordan-frame matter density fluctuation.
The field $\pi$ satisfies a similar relation,
\be
\frac{\nabla^2}{a^2}\pi = - \frac{\frac{\dot f}{4f}}{\frac{3\dot{f}^2}{4f} \tM_{\rm Pl}^2 + c}
\delta \rho
\ee
As a result the Poisson equation for the Jordan-frame metric potential $\Phi$
is modified and becomes
\be
\frac{\nabla^2}{a^2} \Phi = \frac{ \frac{\dot f^2}{f} \tM^2_{\rm Pl} + c }
{\frac{3 \dot f^2}{4f} \tM^2_{\rm Pl} + c} \; \frac{\delta \rho}{2 M_{\rm Pl}^2} ,
\ee
which also writes in term of Newton's potential, $\Psi_{\rm N} = \tilde\Psi_{\rm N}$,
as
\be
\Phi = (1+\epsilon_1) \Psi_{\rm N} \;\;\; \mbox{with} \;\;\;
\epsilon_1 = \frac{\frac{\dot f^2}{4f} \tM^2_{\rm Pl}}{\frac{3\dot{f}^2}{4f} \tM^2_{\rm Pl}+c}
= \frac{2\beta^2}{\bar K'} ,
\ee
and we recover Eq.(\ref{Poisson-K}).
The effect of this modification of the Poisson equation appears naturally in the growth
of the matter density contrast $\delta=\delta\rho/\bar \rho$, which satisfies in the
linear regime
\be
\frac{\partial^2\delta}{\partial t^2} + 2 H \frac{\partial\delta}{\partial t}
-\frac{3}{2} \Omega_{\rm m} H^2 (1+\epsilon_1) \delta = 0 ,
\ee
in agreement with Eq.(\ref{D-linear-Jordan}).
This means that the growth of large-scale structures defines a modified Newton
constant ${\cal G}_{\rm N}^{\rm J}$ as
\be
{\cal G}_{\rm N}^{\rm J} = \frac{ \frac{\dot f^2}{f} \tM^2_{\rm Pl} + c }
{\frac{3 \dot f^2}{4f} \tM^2_{\rm Pl} + c} \;  {\cal G}_{\rm N} = (1+\epsilon_1) \;
{\cal G}_{\rm N} ,
\ee
where ${\cal G}_{\rm N} = 1/8\pi M_{\rm Pl}^2 = 1/8\pi f(\tau) \tM_{\rm Pl}^2$.

Lensing effects are not sensitive to $\Phi$ but to
$\Phi+\Psi= 2 \Psi_{\rm N}$, implying that the lensing potential feels the modified Newton
constant
\be
{\cal G}_{\rm N}^{\gamma} = {\cal G}_{\rm N} .
\ee
It is one of the main features of these models that gravity does not have a unique signature but manifests itself differently for matter and photons.

Similarly, the ratio between the two Jordan-frame Newtonian potentials
$\Psi$ and $\Phi$, defined by $\gamma= \frac{\Psi}{\Phi}$, is not equal to one but
\be
\gamma \equiv \frac{\Psi}{\Phi} = \frac{1-\epsilon_1}{1+\epsilon_1} =
\frac{1-\frac{2\beta^2}{\bar K'}}{1+ \frac{2\beta^2}{\bar K'}} .
\ee
Again we find a deviation from unity which depends on $\frac{2\beta^2}{\bar K'}$. These results coincide with the ones deduced in the main text.

\section{Superluminality and Causality}
\label{sec:causality}

In this appendix, we comment  on the superluminality of scalar perturbations and its link with causality. This issue was analysed for K-essence models in \cite{Babichev:2007dw}. We follow a similar method here.
Let us first expand the K-mouflage action to second order in $\sigma$, where
$\varphi=\bar\varphi + \sigma$. Here $\bar\varphi(\vec x,t)$ is a background configuration
that may depend on scale and time.
We also denote
$\bar\chi=-\bar{g}^{\mu\nu}\pl_{\mu}\bar\varphi \pl_{\nu}\bar\varphi/2\cM^4$,
$\bar K'= K'(\bar \chi)$ and $\bar K''= K''(\bar\chi)$.
The second order part of the action reads
\be
S_{2}= \frac{1}{2}\int d^4x \sqrt{-g}  \left[ -\bar K' \pl^{\mu}\sigma \pl_{\mu}\sigma
+ \frac{\bar K''}{{\cal M}^4} (\pl^{\mu}\bar\varphi \pl_{\mu} \bar\sigma)^2 \right] .
\ee
It is convenient to define the disformal metric
\be
G^{\mu\nu} = \gamma^{-1} \left( \bar K' g^{\mu\nu}
- \frac{\bar K''}{\cM^4} \pl^{\mu}\bar\varphi \pl^{\nu}\bar\varphi \right)
\ee
with
\be
\gamma = (\bar K')^{3/2} (\bar K'+2\bar\chi \bar K'')^{1/2} > 0 .
\label{gamma-def}
\ee
We consider models for which
 $K'>0$ and $K'+2\chi K''>0$ over all $\chi$, so that the metric
$G^{\mu\nu}$ is well defined, see also (\ref{W+-cond}).
Defining the inverse matrix $G_{\mu\nu}$  and the determinant
$G=\det(G_{\mu\nu})$, we have
\be
G= \gamma^2 \, g ,
\label{det-G-munu}
\ee
where $g=\det(g_{\mu\nu})$, and the second-order action can also be written as
\be
S_{2}= -\frac{1}{2} \int d^4 x \, \sqrt{-G} \, G^{\mu\nu} \, \partial_\mu \sigma \partial_\nu \sigma .
\label{S2-Gdisf}
\ee
The disformal metric $G_{\mu\nu}$ is the metric felt by the scalar perturbation and it
is Lorentzian.
Initial-value problems for $\sigma$ are well posed on any smooth
space-like Cauchy surface $\Sigma$ for the metric $G_{\mu\nu}$, and the solution
is unique and propagates causally (see Sec.10 in \cite{Wald1984}).
Around the cosmological background we recover the hyperbolic Klein-Gordon equation
with the propagation speed $c_s$ of Eq.(\ref{cs2-def}).

Space-times equipped with the metric $G_{\mu\nu}$ are stably causal, i.e. there are no
time-like closed loops, provided there exists a globally defined function $f$ on all space-time
which is time-like, i.e. $G^{\mu\nu} \partial_\mu f \partial_\nu f <0$ \cite{Wald1984}.
Following \cite{Babichev:2007dw}, we search for a ``global time'' $f$ that applies
to both geometries $g_{\mu\nu}$ and $G_{\mu\nu}$ and thus implies the
absence of closed causal loops. A simple choice is the cosmic time
$t$, which clearly satisfies the required property for the metric $g_{\mu\nu}$.
Considering the Newtonian gauge, which describes all systems where gravity is weak enough
\be
d s^2 = g_{\mu\nu} d x^\mu d x^\nu = - (1+2\Psi_{\rm N}) d t^2
+ a^2(t) (1-2\Psi_{\rm N}) d \vec x^2 ,
\ee
where $\Psi_{\rm N}$ is the Newtonian potential, we have
\be
g^{\mu\nu} \pl_{\mu} t \pl_{\nu} t = \frac{-1}{1+2\Psi_{\rm N}} < 0
\;\;\; \mbox{for} \;\;\; \Psi_{\rm N}> -1/2 .
\label{CCC-g}
\ee
Since we focus on systems with $|\Psi_{\rm N}| \ll 1$ (e.g., $\Psi_{\rm N} \sim 10^{-6}$
in the Solar System), we have $g^{\mu\nu} \pl_{\mu} t \pl_{\nu} t <0$.
On the other hand, we obtain
\be
G^{\mu\nu} \pl_{\mu} t \pl_{\nu} t = - \frac{\bar K' (1+2\Psi_{\rm N}) +
\bar K'' (\pl_0\bar\varphi)^2/\cM^4}{\gamma (1+2\Psi_{\rm N})^2} ,
\ee
and therefore
\be
G^{\mu\nu} \pl_{\mu} t \pl_{\nu} t < 0 \;\; \mbox{for} \;\;
{\cal C} \equiv \bar K' + \bar K'' \frac{(\pl_0\bar\varphi)^2}{\cM^4} > 0 ,
\label{cC-def}
\ee
where we used the approximation $1+2\Psi_{\rm N} \simeq 1$.
Around the cosmological background, where
$\bar\chi=(d\bar\varphi/d t)^2/2\cM^4$, we obtain
${\cal C}=\bar K'+2\bar\chi\bar K''$, hence ${\cal C}>0$.
Around a static background, we obtain ${\cal C}=\bar K'$ whence ${\cal C}>0$.

For more general backgrounds, we can see from Eq.(\ref{cC-def}) that
${\cal C}>0$ as soon as $\bar K'' \geq 0$.
When  $\bar K'' \leq 0$, we can only conclude in some specific cases. First of all consider cosmological situations where $(\dot{\bar \varphi})^2 \gg (\nabla \bar\varphi)^2$, we have
${\cal C} \sim \bar K'+2\bar\chi\bar K''>0$. Similarly in almost static configurations where $(\dot{\bar \varphi})^2 \ll (\nabla \bar\varphi)^2$, we have
${\cal C}\ge \bar K' +\bar K'' \frac{(\nabla \bar \varphi)^2}{\cM^4}  \sim \bar K'-2\bar\chi\bar K''$. In this almost static limit, where $\bar \chi <0$ and is a large negative number, we must apply the restriction
obtained from the bound on the perihelion advance of the moon, i.e. we require that $\vert K'' \chi \vert \ll K'$ when $\chi \lesssim 1$ \cite{Barreira:2015aea}. This implies that ${\cal C}>0$ for the models passing the solar system tests.
We now examine the intermediate cases when $\bar\chi \sim - (\pl\bar\varphi/\pl t)^2/2\cM^4
\sim - (\nabla\bar\varphi)^2 /2\cM^4$, i.e. the system is both spatially and time varying.
Then, we have
$ {\cal C} \sim \bar K' +2  \bar K'' \bar\chi >0$.
In conclusion, even when $\bar K'' <0$, these three cases cover all of the astrophysical situations of interest, and a  violation of causality is  not present.

Thus, we conclude that $g^{\mu\nu} \pl_{\mu} t \pl_{\nu} t <0$ and
$G^{\mu\nu} \pl_{\mu} t \pl_{\nu} t <0$ and there are no closed causal loops
around usual astrophysical and cosmological backgrounds with $\Psi_{\rm N} > -1/2$.
This analysis fails close to neutron stars or black holes, where $\Psi_{\rm N}$ becomes
large, but this is not related to the K-mouflage model as it already appears
in the metric $g_{\mu\nu}$.
More generally, in such theories where superluminal velocities can be found,
it may be possible to find inhomogeneous and time-dependent backgrounds,
such as boosted bubbles moving apart \cite{Adams2006}, where closed causal loops exist.
Moreover, global properties of space-time may break causality.
However, these two issues also appear within General Relativity itself and one
has to resort to the chronology protection conjecture, which assumes that
the laws of physics prevent the formation of backgrounds leading to closed causal loops.
Then, as argued in \cite{Bruneton2007,Babichev:2007dw}, from the point of view of causality
theories such as K-essence, or K-mouflage in our case, are similar to General Relativity
and do not lead to greater causal paradoxes.
From the perspective of this paper, where we restrict ourselves to linear perturbations
around the cosmological background, we have shown from the analysis above that
causality is preserved, even in the case of model 1 where $c_s>1$ at low redshifts.
Within the effective field theory approach, we only reconstruct the regime $\tilde\chi>0$,
and the regime $\tilde\chi<0$, where problematic backgrounds may appear, is not
reconstructed and beyond the scope of such frameworks.

\end{document}